\documentclass[aps,prd,twocolumn,superscriptaddress,nofootinbib,longbibliography]{revtex4-1}
\usepackage{amsmath,amssymb, amsthm,amstext}
\usepackage{graphicx}
\usepackage{color}
\usepackage{array, enumerate}
\usepackage{bm}
\usepackage[breaklinks,colorlinks,citecolor=blue]{hyperref}
\usepackage{braket}
\usepackage{txfonts}
\usepackage[normalem]{ulem}

\def\be{\begin{equation}}
\def\ee{\end{equation}}

\begin{document}

\title{Waveforms and Fluxes of Generic Extreme-Mass-Ratio Inspirals with a Spinning Secondary}
\author{Qiuxin Cui}
\email{cuiqiuxin@shao.ac.cn}
\affiliation{State Key Laboratory of Radio Astronomy and Technology, Shanghai Astronomical Observatory, Chinese Academy of Sciences, 80 Nandan Road, Shanghai 200030, China}
\affiliation{School of Astronomy and Space Science, University of Chinese Academy of Sciences, Beijing, 100049, People’s Republic of China}
\author{Wen-Biao Han }
\email[Contact author:]{wbhan@shao.ac.cn}
\affiliation{State Key Laboratory of Radio Astronomy and Technology, Shanghai Astronomical Observatory, Chinese Academy of Sciences, 80 Nandan Road, Shanghai 200030, China}
\affiliation{School of Fundamental Physics and Mathematical Sciences, Hangzhou Institute for Advanced Study, UCAS, Hangzhou 310024, People’s Republic of China}
\affiliation{School of Astronomy and Space Science, University of Chinese Academy of Sciences, Beijing, 100049, People’s Republic of China}
\begin{abstract}
Extreme mass-ratio inspirals (EMRIs), comprising a stellar-mass compact object (CO) orbiting a supermassive black hole (BH), are key targets for future space-based gravitational-wave (GW) observatories. Incorporating the spin of the secondary body into waveform models not only enhances measurement precision but also offers insight into the spin distribution of stellar-mass COs. In this work, we construct the flux and waveform for an EMRI with a spinning secondary in a Kerr background under the linear-spin approximation. Using the radiative prescription (half-retarded minus half-advanced field), we calculate orbit-averaged fluxes for the fundamental constants of motion, including the energy, angular momentum, and the Carter-like constant. This framework provides a tractable route to generating waveforms that incorporate the secondary spin.
\end{abstract}
\date{\today}
\maketitle

\section{Introduction}
Since the first direct detection of a gravitational wave (GW) signal from a binary black hole merger in 2015 \cite{LIGO2016}, the LIGO-Virgo-KAGRA (LVK) collaboration has announced more than 100 compact binary merger events \cite{LIGO2023}. The gravitational waves from these stellar-mass compact binaries at coalescence are in a relatively high frequency band. The future space-borne gravitational wave detectors, such as LISA \cite{LISA2017}, Taiji \cite{Taiji2020}, and Tian-Qin \cite{TianQin2016}, are most sensitive in the milli-Hertz (mHz) band. One of the primary sources for these space-borne interferometers is extreme mass-ratio inspirals (EMRI) \cite{LISA_EMRI2017}, which consists of a stellar-mass compact object, e.g., a stellar-mass black hole (sBH) with mass $\mathcal{O}(1-100)$ $M_\odot$, and a supermassive black hole (SMBH).

It relies on the matched-filter technology to dig out the EMRI event in the continuous long signal or resolve system parameters, which requires precise system simulation. Due to the extreme mass ratio $\epsilon$, we can apply the two-timescale analysis to the EMRI system \cite{Hinderer2008}. Generally, angle variables $q_{\alpha}$ evolve on the orbit timescale $\sim 2\pi /\omega_{\alpha}$ as
\begin{equation}
\frac{dq_{\alpha}}{d\tau} = \omega_{\alpha}^{(0)}(Q_{\lambda}) + \epsilon \omega^{(1)}_{\alpha}(Q_{\lambda},q_i) + \mathcal{O}(\epsilon^2),
\end{equation}
where $\omega_{\alpha}$ are the orbit frequencies and $Q_{\lambda}$ are the conserved quantities of geodesic motion. While $Q_{\lambda}$ evolve on the radiation timescale $\sim 2\pi/\epsilon\omega$ as
\begin{equation}
\frac{dQ_{\lambda}}{d\tau} = \epsilon G^{(1)}_{\lambda}(Q_{\sigma},q_i) + \epsilon^2 G^{(2)}_{\lambda}(Q_{\sigma},q_i) + \mathcal{O}(\epsilon^3).
\end{equation}
Since the detected signal will persist for about one year, e.g., $\sim 1/\epsilon$ cycles for the orbit motion, the leading-order simulation named adiabatic approximation (0PA) should involve $\omega^{(0)}_{\alpha}$ and $\epsilon G^{(1)}_{\lambda}$. And the subleading-order effects are $\epsilon \omega^{(1)}_{\alpha}$ and $\epsilon^2G^{(2)}_{\lambda}$, named the first post-adiabatic approximation (1PA). By Lindblom's criteria \cite{Lindblom2008}, the 1PA simulation should fulfill the precision requirement of future detection. 

Ignoring the perturbation of the background metric, the motion of the test body is described by the Dixon laws \cite{Dixon1970,Dixon1974}. For compact objects (such as the BH), the zero-order term is the geodesic motion, which contributes to $\mathcal{O}(\epsilon^0)$, and the spin of the compact object ($s$, named the secondary spin, as the spin of SMBH, $a$, is named the primary spin) contributes to at $\mathcal{O}(\epsilon)$. Other higher-pole moments should contribute to $\mathcal{O}(\epsilon^2)$. When it comes to the gravitational perturbation induced by the secondary body (i.e., the self-force effect), Ref.~\cite{Harte2012} showed that the motion still follows the same equation as a test body, but in an effective metric instead of the background metric. Thus, considering the extreme mass ratio $\epsilon$, the geodesic motion should contribute to the $\mathcal{O}(\epsilon)$ self-force effect, including the dissipative part and the conservative one. While the secondary spin should contribute to the $\mathcal{O}(\epsilon^2)$ self-force effect. For the 1PA approximation, we should take into account all the $\mathcal{O}(\epsilon)$ self-force effects and the dissipative part of the $\mathcal{O}(\epsilon^2)$ one, so other higher-pole moments can be safely discarded.

In this paper, we concentrate on the effect of the secondary spin instead of the full 1PA effect, i.e., only the dissipative self-force effects are taken into account. Specifically speaking, although $s/M \sim \epsilon \ll 1$, we will handle these two approximations separately and take their linear approximations respectively. Practically speaking, we compute the linear metric perturbation induced by the orbit of the Mathison-Papapetrous-Dixon (MPD) equation \cite{mathisson1937neue,Papapetrou1951} (which is the dipole approximation of the Dixon law) with the linear-spin approximation, and then make use of this to compute the inspiral, i.e., orbit-averaged evolution of the constants of motion. This approach can be regarded as an adiabatic prescription of a spinning particle.

The secondary spin is an interesting and significant component in the EMRI system, not only because it is necessary for the 1PA simulation but also because it can reveal certain important characteristics of the EMRI system. As Ref.~\cite{Cui2025} showed, the secondary spin can be used to distinguish dry (loss cone) EMRIs (where the secondary BHs were born in the collapse of individual massive stars and are of low spin) and Hills EMRIs (where the secondary BHs are remnants of massive star binaries and the secondary spins follow a bimodal distribution). Besides, Ref.~\cite{Piovano2020,XuLujia2025} suggested that the secondary spin may also be used to distinguish the stellar-mass BH from other kinds of compact objects. This is why we first focus our attention on the secondary spin in the 1PA effect.

For a spinning secondary, some special orbits have been carefully studied, such as the straight falling one in Ref.~\cite{Mino1996}, the circular-aligned one in Ref.~\cite{Tanaka1996,Han2010,Piovano2020II,LG2017,Harms2016}, the eccentric equatorial one in Ref.~\cite{Viktor2021,Viktor2022}, the Schwarzschild one in Ref.~\cite{Mathews2022II,Viktor2024}, and the quasi-spherical one in Ref.~\cite{Viktor2025}. Although some research \cite{Gair2011,Piovano2021} argued that the secondary spin is unmeasurable for LISA with a typical mass ratio $\sim 10^{-5}$, the motion of the secondary is strictly constrained to an equatorial and further circular orbit in their discussion. Recently, Ref.~\cite{Viktor2025} found that with a certain inclination, neglecting the secondary spin will induce a large mismatch. For generic orbits, the evolution of the energy and angular momentum has been calculated by the flux-balance law in Ref.~\cite{Viktor2023,Piovano2025}, but this is not sufficient to evolve the generic MPD orbit, which is precisely what this research aims to achieve. 

A few days before the release of this work, an article \cite{Viktor2026} appeared that similarly computed the flux for the Carter-like constant, based on the flux-balance law in Ref.~\cite{Josh2025,Grant2025,Witzany2025}. In contrast, the derivation in this work basically follows the method in Ref.~\cite{Sago2006}, i.e., starting from the radiative field (half-retarded minus half-advanced field).

This paper is organized as follows. In Sec.~\ref{Sec:OrbitMotion}, we review the analytical solution of geodesic orbits in a Kerr spacetime, then we turn to the MPD equation and introduce the orbital formulation we adopt under the linear-spin approximation. In Sec.~\ref{sec:MetricPerturbation}, we present the computation of the asymptotic wave and the radiative field, based on the Teukolsky formalism. In Sec.~\ref{Sec:OrbitInspiral}, we derive the orbit-averaged evolution equation of the constants of motion, and show how to incorporate the radiative field into the flux calculation. In Sec.~\ref{Sec:Result}, we perform several checks for the computation of our code, and then present an example of inspiral waveforms with a spinning secondary. Throughout this paper, the geometrical units with $c=G=1$ and the Einstein summation convention are adopted.
\section{Orbit Motion}\label{Sec:OrbitMotion}
\subsection{Geodesic orbit in Kerr spacetime}
The Kerr metric in the Boyer-Lindquist coordinate $\{t,r,z=\cos\theta,\varphi\}$ is 
\begin{equation}
\begin{aligned}
ds^2 = &-\left(1-\frac{2Mr}{\Sigma}\right)dt^2 + \frac{\Sigma}{\Delta}dr^2 + \frac{\Sigma}{1-z^2} dz^2 \\
&+\left[\left(r^2+a^2\right)(1-z^2)+\frac{2Mra^2(1-z^2)^2}{\Sigma}\right]d\varphi^2 \\
&- \frac{4Mra(1-z^2)}{\Sigma} dt d\varphi, 
\end{aligned}
\end{equation}
where $\Delta := r^2 - 2Mr + a^2$ and $\Sigma := r^2 + a^2z^2$. The Hamilton-Jacobi equation for a test particle is separable \cite{Carter1968}, leading to the following equations of motion (EOM)
\begin{equation}
\begin{aligned}
\left(\frac{dr}{d\lambda}\right)^2 &= R(r), \\
\left(\frac{dz}{d\lambda}\right)^2 &= Z(z), \\
\frac{dt}{d\lambda} &= \left(\Sigma+\frac{2Mr\left(r^2+a^2\right)}{\Delta}\right)E-\frac{2Mra}{\Delta}J_z,\\
\frac{d\phi}{d\lambda} &= \frac{2Mra}{\Delta}E + \frac{\Sigma-2Mr}{\Delta(1-z^2)}J_z,
\end{aligned}
\end{equation}
where the Mino time $\lambda$ is defined as $\lambda := \int d\tau/\Sigma$, and the radial function and the polar function are
\begin{equation}
\begin{aligned}
R(r) &:= \left[\left(r^2 + a^2\right)E - aJ_z\right]^2-\Delta \left(K + r^2\right),\\
Z(z) &:= \left(1-z^2\right)\left(K-a^2z^2\right) - \left[J_z - aE(1-z^2)\right]^2.
\end{aligned}
\end{equation}
$\{E,J_z,K\}$ are those constants used to separate the EOM, and they can also be derived using the symmetric properties of Kerr spacetime as in the form
\begin{equation}
\begin{aligned}
E &\equiv -\xi^{\mu}_{(t)}u_{\mu}, \\
J_z &\equiv \xi^{\mu}_{(\varphi)}u_{\mu}, \\
K &\equiv K_{\mu\nu}u^{\mu}u^{\nu},
\end{aligned}
\end{equation}
where $\{\xi^{\mu}_{(t)},\xi^{\mu}_{(\varphi)},K_{\mu\nu}\}$ are the famous Killing vectors/tensors that Kerr spacetime admits. 
For bound orbits, $r$ ranges between $\{r_a,r_p\}$ which are two of the four roots in $R(r)$ and $z$ ranges between $\{-z_-,z_-\}$ where $z_-$ is one of the two roots in $Z(z)$. 
Therefore, parameterizing the bound orbit directly using these roots, rather than the separation constants, is more straightforward. A common choice is 
\begin{equation}
\begin{aligned}
\text{semi-latus rectum:}\ \ &p := \frac{2r_ar_p}{r_a + r_p}, \\
\text{eccentricity:}\ \ &e:= \frac{r_a-r_p}{r_a + r_p}, \\
\text{inclination:}\ \ &\theta_{\text{min}} := \arccos z_-.
\end{aligned}
\end{equation}
An elegant transformation between the root parameters and the constant parameters is given in \cite{Schmidt2002}. The fundamental periods $\Lambda$ and frequencies $\Upsilon$ with respect to $\lambda$ are defined as
\begin{equation}
\begin{aligned}
\Lambda_r = 2\int_{r_a}^{r_p}\frac{dr}{\sqrt{R(r)}} \ \ &, \ \ \Lambda_z = 4\int_0^{z_-}\frac{dz}{\sqrt{Z(z)}} \\
\Upsilon_r := \frac{2\pi}{\Lambda_r} \ \ &, \ \ \Upsilon_z := \frac{2\pi}{\Lambda_z}
\end{aligned}
\end{equation}
Due to the complete separability in the radial and polar directions, these geodesics can be expressed analytically using elliptic functions \cite{Fujita2009}. In particular, the expressions for the coordinate time $t(\lambda)$ and the azimuthal angle $\varphi(\lambda)$ each consist of three parts: the linearly increasing term, the radial oscillatory term, and the polar oscillatory term,
\begin{equation}
\begin{aligned}
t(\lambda) &= \Gamma\lambda + t^{(r)}(\lambda) + t^{(z)}(\lambda), \\
\varphi (\lambda) &= \Upsilon_{\varphi}\lambda + \varphi^{(r)}(\lambda) + \varphi^{(z)}(\lambda).
\end{aligned}
\end{equation}

\subsection{MPD orbit in Kerr spacetime}\label{Sec:MPDOrbit}
In general relativity, the motion of a classically spinning particle is described by MPD equations as below \cite{Dixon1970,Dixon1974}, 
\begin{equation}\label{eq:MPDEQ}
\begin{aligned}
\frac{DP^{\mu}}{d\tau} &= -\frac{1}{2}R^{\mu}_{\ \ \nu\rho\sigma}v^{\nu}S^{\rho\sigma}, \\
\frac{DS^{\mu\nu}}{d\tau} &= 2P^{[\mu}v^{\nu]},
\end{aligned}
\end{equation}
where $\{v^{\mu},P^{\mu},S^{\mu\nu}\}$ represent the four-velocity, the momentum, and the spin tensor. Their magnitudes can be defined as
\begin{equation}
\begin{aligned}
\mathfrak{m}_s &:= -P_{\mu}v^{\mu}, \\
S &:= \sqrt{\frac{1}{2}S^{\mu\nu}S_{\mu\nu}}, \\
m_s &:= \sqrt{-P^{\mu}P_{\mu}}.
\end{aligned}
\end{equation}
Since $v^{\mu}$ and $P^{\mu}$ are generally not aligned, the two mass parameters $m_s$ and $\mathfrak{m}_s$ are also not equal. If the background spacetime admits a symmetry described by the Killing vector field $\xi^{\mu}$, the covariant conservation of energy-momentum $\nabla_{\alpha}T^{\alpha\beta}\equiv 0$ will induce a constant of motion as
\begin{equation}
C = P^{\mu}\xi_{\mu} - \frac{1}{2}S^{\mu\nu}\xi_{[\mu ; \nu]}
\end{equation}
Thus, in Kerr spacetime, there exist two constants of motion analogous to the energy $E$ and angular momentum $J_z$ in the spinless case:
\begin{equation}\label{eq:EnCandJzC}
\begin{aligned}
E&:=-\frac{1}{m_s}\xi^{(t)}_{\mu}P^{\mu} + \frac{1}{2m_s}S^{\mu\nu}\xi_{\mu ; \nu} \\
J_z&:=\frac{1}{m_s}\xi^{(\varphi)}_{\mu}P^{\mu} - \frac{1}{2m_s}S^{\mu\nu}\xi_{\mu ; \nu}
\end{aligned}
\end{equation}

In order to fix the (center of mass) frame where these quantities are defined, we adopt the Tulczyjew-Dixon (TD) condition $S^{\mu\nu}P_{\nu}=0$ in this paper. Thus, with this condition, the four-velocity can be expressed as \cite{Semerak1999,Semerak2007}
\begin{equation}\label{v-p-relation}
v^{\mu} = \frac{\mathfrak{m}_s}{m_s^2}\left(P^{\mu} + \frac{2S^{\mu\nu}R_{\nu\gamma\kappa\lambda}P^{\gamma}S^{\kappa\lambda}}{4m_s^2 + R_{\mu\nu\kappa\lambda}S^{\mu\nu}S^{\kappa\lambda}}\right),
\end{equation}
and it is easy to show that $\dot{S}=0$, $\dot{m}_s=0$, while $\dot{\mathfrak{m}_s}\neq 0$. Another quantity, the spin vector, is defined as
\footnote{In this work, the contravariant Levi-Civita tensor is defined as $\epsilon^{\mu\nu\kappa\lambda}:=\sqrt{|\text{det}(g)|}\ \varepsilon^{\mu\nu\kappa\lambda}$, where $\varepsilon^{\mu\nu\kappa\lambda}$ is the Levi-Civita symbol.}
\begin{equation}
S^{\mu} := -\frac{1}{2m_s}\epsilon^{\mu\nu\kappa\lambda}P_{\nu}S_{\kappa\lambda}.
\end{equation}
It is totally equivalent to the spin tensor,
\begin{equation}
\begin{aligned}
S^{\mu\nu} &= \epsilon^{\mu\nu\kappa\lambda}P_{\kappa}S_{\lambda}, \\
S &= \sqrt{S^{\mu}S_{\mu}}.
\end{aligned}
\end{equation}

In EMRI systems where $\mathfrak{m}_s \sim m_s \ll M$, the spin magnitude satisfies $S \leq m^2_s \ll m_sM$. Thus, no matter in the equations of motion as
\begin{equation}
\begin{aligned}
\frac{dP^{\mu}}{d\tau}&=-\Gamma^{\mu}_{\rho\sigma}v^{\rho}P^{\sigma}-\frac{1}{2}R^{\mu}_{\ \ \nu\rho\sigma}v^{\nu}S^{\rho\sigma} \\
&\approx \mathcal{O}(\frac{m_s}{M}) + \mathcal{O}(\frac{m_s^2}{M^2}),
\end{aligned}
\end{equation}
or in the stress-energy tensor as
\begin{equation}
\begin{aligned}
T^{\mu\nu}(x^{\mu})&= \int d\tau \bigg[\frac{\delta^{(4)}(x^{\mu}-z^{\mu}(\tau))}{\sqrt{-g}}P^{(\mu}v^{\nu)} \\
&\hspace{1cm} -\nabla_{\sigma}\left(\frac{\delta^{(4)}(x^{\mu}-z^{\mu}(\tau))}{\sqrt{-g}}S^{\sigma(\mu}v^{\nu)}\right) + \mathcal{O}(S^2)\bigg] \\
&=\int d\tau \ \ \mathcal{O}(m_s) + \mathcal{O}(m_s\frac{m_s}{M}) + \mathcal{O}(m_s\frac{m_s^2}{M^2}),
\end{aligned}
\end{equation}
the contribution from the spin part is a higher-order term compared with the leading one. So we adopt the linear-spin approximation in this paper and ignore the higher-order terms. We define the specific spin tensor as $s^{\mu\nu}=S^{\mu\nu}/m_s$, and so does the spin vector. The specific spin magnitude $s=S/m_s$ thus satisfies $s\sim m_s\ll M$. This approximation leads to several significant simplifications. From Eq.~\ref{v-p-relation}, the four-velocity would be aligned with the momentum, and thus $m_s$ would be equal to $\mathfrak{m}_s$. Consequently, $P^{\mu}=m_sv^{\mu}=\mathfrak{m}_sv^{\mu}$. The EOM would also be simplified as
\begin{equation}\label{MPDequationL}
\begin{aligned}
\frac{Dv^{\mu}}{d\tau} &= -\frac{1}{2}R^{\mu}_{\ \ \nu\rho\sigma}v^{\nu}s^{\rho\sigma}, \\
\frac{Ds^{\mu\nu}}{d\tau} &=0 \ \ \text{or} \ \ \frac{Ds^{\mu}}{d\tau} = 0 .
\end{aligned}
\end{equation}
Under the linear-spin approximation, the Hamilton-Jacobi equation for a spinning particle remains separable \cite{Vojtech2019}, yielding two additional constants of motion, which were also derived by \cite{Rudiger1983},
\begin{equation}\label{eq:CarterCandSpinC}
\begin{aligned}
K &:= K_{\mu\nu}v^{\mu}v^{\nu} - 2v^{\mu}s^{\rho\sigma}\left(Y_{\mu\rho ; \kappa}Y^{\kappa}_{\ \ \sigma} + Y_{\rho\sigma ; \kappa}Y^{\kappa}_{\ \ \mu}\right), \\
s_{||} &:= \frac{Y_{\mu\nu}v^{\mu}s^{\nu}}{\sqrt{K_{\mu\nu}v^{\mu}v^{\nu}}},
\end{aligned}
\end{equation}
where $Y_{\mu\nu}$ is the Yano-Killing tensor, $K$ corresponds to the Carter constants in the spinless case, and $s_{||}$ represents the projection of the specific spin vector onto the orbit angular momentum $Y_{\mu\nu}v^{\mu}$.
From Eq.~\ref{MPDequationL}, the spin vector is evolving through parallel transport along the worldline. Furthermore, since only the linear spin terms are retained, this parallel transport can be approximated as occurring along a geodesic—specifically, one that shares the same values of $\{E,J_z,K\}$ as the MPD orbit—which deviates from the physical worldline only at $\mathcal{O}(s/M)$. The expression of the parallel transport tetrad for geodesic was derived in \cite{Marck1983,vandeMeent2020}.
\begin{equation}
\begin{aligned}
e^0_{\mu} &:= \left(-E,\frac{1}{\Delta}\frac{dr}{d\lambda},\frac{1}{1-z^2}\frac{dz}{d\lambda},J_z\right), \\
e^1_{\mu} &:= \cos\psi(\lambda)\tilde{e}^1_{\mu} + \sin\psi(\lambda)\tilde{e}^2_{\mu}, \\
e^2_{\mu} &:= -\sin\psi(\lambda)\tilde{e}^1_{\mu} + \cos\psi(\lambda)\tilde{e}^2_{\mu}, \\
e^3_{\mu} &:= \left(
\begin{array}{ll}
a\frac{rdz/d\lambda+zdr/d\lambda}{\Sigma\sqrt{K}} \\
az\frac{(r^2+a^2)E-aJ_z}{\Delta\sqrt{K}} \\
\frac{rJ_z-aE(1-z^2)}{(1-z^2)\sqrt{K}} \\
-\frac{a^2z(1-z^2)dr/d\lambda+r(r^2+a^2)dz/d\lambda}{\Sigma\sqrt{K}}
\end{array}\right),
\end{aligned}
\end{equation}
where $\tilde{e}^1_{\mu}$ and $\tilde{e}^2_{\mu}$ are defined as
\begin{equation}
\begin{aligned}
\tilde{e}^1_{\mu} := &\left(
\begin{array}{ll}
\frac{-\Xi rdr/d\lambda+\frac{a^2z}{\Xi}dz/d\lambda}{\sqrt{K}\Sigma} \\
\Xi r \frac{(r^2+a^2)E-aJ_z}{\sqrt{K}\Delta} \\
-\frac{az}{\sqrt{K}\Xi}\left(aE-\frac{J_z}{1-z^2}\right) \\
a\frac{\Xi^2 r (1-z^2)dr/d\lambda - z(r^2+a^2)dz/d\lambda}{\sqrt{K}\Xi \Sigma} 
\end{array}
\right),  \\ 
\tilde{e}^2_{\mu} :=& \left(
\begin{array}{ll}
-\frac{E}{\Xi}+\frac{(1-\Xi^2)\left[(r^2+a^2)E - aJ_z\right]}{\Xi \Sigma} \\
\frac{\Xi}{\Delta} \frac{dr}{d\lambda} \\
\frac{1}{\Xi (1-z^2)}\frac{dz}{d\lambda} \\
\Xi J_z + \frac{(1-\Xi^2)(r^2+a^2)\left[J_z - a(1-z^2)E\right]}{\Xi \Sigma}
\end{array}
\right),
\end{aligned}
\end{equation}
and $\Xi := \sqrt{\frac{K - a^2 z^2}{K + r^2}}$. The precession angle $\psi(\lambda)$ evolves as
\begin{equation}
\frac{d\psi}{d\lambda} = \sqrt{K} \left[\frac{(r^2+a^2)E-aJ_z}{K+r^2} + a\frac{J_z-a(1-z^2)E}{K-a^2z^2}\right].
\end{equation}
Similar to the coordinate time and the azimuthal angle, the precession angle has an analytical form as
\begin{equation}
\psi(\lambda) = \Upsilon_{\psi}\lambda + \psi^{(r)}(\lambda) + \psi^{(z)}(\lambda).
\end{equation}
Thus, the spin vector could be projected into this tetrad with the projection components being constants as 
\begin{equation}
s^{\mu} = s_{||}e_{3}^{\mu} + s_{\perp}e_{1}^{\mu} + \mathcal{O}\left(s^2/M^2\right).
\end{equation}

In \cite{Skoupy2025}, the author provides an elegant method to resolve the orbit motion $x^{\mu}$ analytically. The key idea is to map the physical worldline to a virtual worldline via the transformation $\tilde{x}^{\mu}:=x^{\mu}+\delta x^{\mu}(r,z,\psi)$. The virtual worldline $\tilde{x}^{\mu}$ can can be associated with a virtual geodesic, which has an analytic solution. So inversely, the physical worldline has the analytical form as
\begin{equation}\label{eq:MPDorbitEQ}
\begin{aligned}
t(\tilde{\lambda};C) &= \tilde{t} - \delta t(\tilde{r},\tilde{z},\psi) \\
&= t_{\text{g}}(\tilde{\lambda};\tilde{C}) - \frac{3s_{||}}{2\sqrt{K}}\tau_{\text{g}}(\tilde{\lambda};\tilde{C}) - \delta t(\tilde{\lambda}), \\
x^{k}(\tilde{\lambda};C) &= \tilde{x}^k - \delta x^k(\tilde{r},\tilde{z},\psi) \\
&= x^k_{\text{g}}(\tilde{\lambda};\tilde{C}) - \delta x^k(\tilde{\lambda}),
\end{aligned}
\end{equation}
where the subscript “$\text{g}$” refers to the geodesic expression, and the auxiliary quantities $\tilde{C}=\{\tilde{E},\tilde{J}_z,\tilde{K}\}$ are defined as 
\begin{equation}
\begin{aligned}
\tilde{E} &:= E + s_{||}\frac{1-E^2}{2\sqrt{K}}, \\
\tilde{J}_z &:= J_z + s_{||}\frac{a-J_zE/2}{\sqrt{K}}, \\
\tilde{K} &:= K + s_{||}\frac{3a(J_z-aE)-KE}{\sqrt{K}},
\end{aligned}
\end{equation}
and the deformed Mino parameter $\tilde{\lambda}$ satisfies
\begin{equation}
\frac{d\tau}{d\tilde{\lambda}} = \left(1-\frac{3s_{||}E}{2\sqrt{K}}\right)\left(\tilde{r}^2+a^2\tilde{z}^2\right).
\end{equation}
It should be noted that, as is stressed by \cite{Skoupy2025}, the $t_\text{g}(\tilde{\lambda};\tilde{C})$ and $x^k_\text{g}(\tilde{\lambda};\tilde{C})$ in Eq.~\ref{eq:MPDorbitEQ} should be understood purely as functional expressions instead of any particular worldline.

From Eq.~\ref{eq:MPDorbitEQ}, we can see that $r$ or $z$ is no longer strictly periodic now, since both $r$ and $z$ appear in the expression of $\delta x^{\mu}(r,z,\psi)$. But the expression of $\tilde{t}$ and $\tilde{x}^k$ still admits periodicity with the time parameter being $\tilde{\lambda}$, and we will use the tilde symbol to label its frequency as $\tilde{\Upsilon}$ and so do other quantities. For example,
\begin{equation}
\tilde{\Gamma} = \Gamma_{\text{g}}(\tilde{C}) - \frac{3s_{||}}{2\sqrt{K}}\Upsilon_{\tau,\text{g}}(\tilde{C}).
\end{equation}

With the spin vector $s^{\mu}$ and the coordinate $x^{\mu}$ both known, the four-velocity $v^{\mu}$ can also be recovered from Eq.~\ref{eq:EnCandJzC} and Eq.~\ref{eq:CarterCandSpinC}, combined with the normalization condition. Having established the analytical description of the spinning secondary’s orbit in the background Kerr spacetime, we now proceed to compute the gravitational perturbations it generates.

\section{Metric Perturbation}\label{sec:MetricPerturbation}
\subsection{Teukolsky formalism}
The perturbation of the gravitational field in Kerr spacetime is described by the famous Teukolsky equation in the form \cite{Teukolsky1973,Sago2006}
\begin{equation}
\begin{aligned}
&\left[\frac{\left(r^2 + a^2\right)^2}{\Delta} - a^2\sin^2\theta\right] \frac{\partial^2 {}_{-2}\Psi}{\partial t^2} + \frac{4Mar}{\Delta} \frac{\partial^2 {}_{-2}\Psi}{\partial t \partial \varphi} \\
&+\left(\frac{a^2}{\Delta} - \frac{1}{\sin^2\theta}\right) \frac{\partial^2 {}_{-2}\Psi}{\partial \varphi^2}-\Delta^{2}\frac{\partial}{\partial r}\left(\Delta^{-1}\frac{\partial {}_{-2}\Psi}{\partial r}\right) \\
&- \frac{1}{\sin \theta} \frac{\partial}{\partial \theta}\left(\sin \theta \frac{\partial {}_{-2}\Psi}{\partial \theta}\right)+4\left[\frac{a\left(r - M\right)}{\Delta} + i\frac{\cos\theta}{\sin^2\theta}\right]\frac{\partial{}_{-2}\Psi}{\partial \varphi} \\
&+4\left(\frac{M\left(r^2 - a^2\right)}{\Delta} - r - ia\cos\theta\right)\frac{\partial {}_{-2}\Psi}{\partial t} + \left(4\cot^2\theta + 2\right){}_{-2}\Psi \\
&= -4\pi \Sigma \mathcal{T},
\end{aligned}
\end{equation}
where ${}_{-2}\Psi := \rho^4\Psi_4$, $\rho:=r-iaz$, and $\Psi_4$ is the Weyl curvature scalar in the Newman-Penrose Formalism where the null tetrad is defined as $\lambda^{\mu}_a=\left\{l^{\mu},n^{\mu},m^{\mu},\overline{m}^{\mu}\right\}$ with
\begin{equation}
\begin{aligned}
l^{\mu} &:= \left(\frac{r^2+a^2}{\Delta},1,0,\frac{a}{\Delta}\right), \\
n^{\mu} &:= \frac{1}{2\Sigma}\left(r^2+a^2,-\Delta,0,a\right), \\
m^{\mu} &:= \frac{\sqrt{1-z^2}}{\sqrt{2}\overline{\rho}}\left(ia,0,-1,\frac{i}{1-z^2}\right).
\end{aligned}
\end{equation}
$\mathcal{T} := {}_{-2}\tau_{\mu\nu} [T^{\mu\nu}]$ is the source term with the operator ${}_{-2}\tau_{\mu\nu}$ defined as
\begin{equation}
\begin{aligned}
{}_{-2}\tau_{\mu\nu} :=& -\frac{\Delta}{\rho^4\overline{\rho}}\Bigg[\left(\mathcal{L}_{-1}\frac{\rho^4}{\overline{\rho}^2}\mathcal{D}^{\dagger}_{-1} + \mathcal{D}^{\dagger}_{-1}\frac{\rho^4}{\overline{\rho}^2}\mathcal{L}_{-1} \right)\frac{\Sigma^2\left(n_{\mu}\overline{m}_{\nu} + \overline{m}_{\mu}n_{\nu}\right)}{2\sqrt{2}} \\
&\ \ \ \ +\frac{1}{\Delta}\mathcal{L}_{-1}\rho^4\mathcal{L}_0\Sigma\rho n_{\mu}n_{\nu} + \frac{\Delta}{2}\mathcal{D}^{\dagger}_0 \rho^4\mathcal{D}^{\dagger}_0\frac{\rho^2}{\overline{\rho}}\overline{m}_{\mu}\overline{m}_{\nu}  \Bigg],
\end{aligned}
\end{equation}
where 
\begin{equation}
\begin{aligned}
\mathcal{D}_n &:= \partial_r + \frac{r^2 + a^2}{\Delta}\partial_t + \frac{a}{\Delta}\partial_{\varphi} + \frac{2n(r-M)}{\Delta}, \\
\mathcal{L}_n &:= \partial_{\theta} - \frac{i}{\sin\theta}\partial_{\varphi} - ia\sin\theta\partial_t + n\cot\theta,
\end{aligned}
\end{equation}
and the superscript $\dagger$ refers to $(\partial_t,\partial_{\varphi}) \rightarrow (-\partial_t,-\partial_{\varphi})$.

Applying a Fourier decomposition to the Teukolsky equation with the separation of the radial and polar variables
\begin{equation}
{}_{-2}\Psi = \sum_{m}\int dw \ e^{-iwt+im\varphi}S_{mw}(z)R_{mw}(r),
\end{equation}
the partial differential equation can be divided into two ordinary differential equations. For the polar variable
\begin{equation}
\begin{aligned}
\frac{d}{dz}&\left[(1-z^2)\frac{d}{dz}S_{mw}(z)\right] + \bigg[(awz)^2 -(aw)^2+ 4awz - 2  \\
&+ 2maw - \frac{(m-2z)^2}{1-z^2}+\lambda_{mw}\bigg]S_{mw}(z) = 0 ,
\end{aligned}
\end{equation}
where $\lambda_{mw}$ is the constant introduced when separating the equation. The solutions of the above equation are known as spin-weighted spheroidal harmonics with spin weight $-2$, i.e. $S^{aw}_{lm}(z)$ where the degree integer $l=\text{max}[2,|m|]$, the normalization is defined as
\begin{equation}
\int_{-1}^{1}\vert S^{aw}_{lm}\vert^2 dz = 1,
\end{equation}
and $\lambda_{mw}$ reduce to the discrete eigenvalue $\lambda_{lmw}$. With these bias functions, the solution of ${}_{-2}\Psi$ can be further divided into
\begin{equation}
{}_{-2}\Psi = \sum_{lm} \int dw \ e^{-iwt + im\varphi} \frac{S^{aw}_{lm}(z)}{\sqrt{2\pi}} R_{lmw}(r).
\end{equation}
Thus, the radial equation is
\begin{equation}\label{eq:RadialEq}
\Delta^2 \frac{d}{dr}\left(\frac{1}{\Delta}\frac{dR_{lmw}}{dr}\right) - V(r)R_{lmw} = T_{lmw},
\end{equation}
where
\begin{equation}
\begin{aligned}
V(r) &:= \lambda_{lmw} + 8iwr - \frac{\mathcal{K}^2 + 4i(r-M)\mathcal{K}}{\Delta}, \\
\mathcal{K} &:= (r^2 + a^2)w - ma,
\end{aligned}
\end{equation}
and the source term $T_{lmw}$ is
\begin{equation}
\begin{aligned}
T_{lmw}(r) :=& \int d\Omega dt \ e^{iwt-im\varphi} \left( 2\Sigma\mathcal{T}\right)\frac{S^{aw}_{lm}(z)}{\sqrt{2\pi}}, \\
4\pi\Sigma\mathcal{T} =& \int dw\sum_{lm}T_{lmw}(r)e^{im\varphi-iwt}\frac{S^{aw}_{lm}(z)}{\sqrt{2\pi}} .
\end{aligned}
\end{equation}
Submitting the expression of $\mathcal{T}$, the source term can be organized into a more concise form \cite{Skoupy2023}
\begin{equation}
T_{lmw} = \int dtd\Omega \ e^{iwt-im\varphi}\frac{\Delta^2}{\sin\theta}\sum_{ab}\sum_{i=0}^{N_{ab}}\frac{\partial^i}{\partial r^i}\left(f^{(i)}_{ab}\sqrt{-g}\ T_{ab}\right),
\end{equation}
where $ab \in \{nn,n\overline{m},\overline{mm}\}$, $\{N_{nn}=0,N_{n\overline{m}}=1,N_{\overline{mm}}=2\}$ and
\begin{equation}
\begin{aligned}
f^{(0)}_{nn} &= -\frac{2\rho^2}{\Delta^2}\left(L_1L_2 - 2ia\rho^{-1}\sqrt{1-z^2}L_2\right)\frac{S^{aw}_{lm}}{\sqrt{2\pi}}, \\
f^{(0)}_{n\overline{m}} &= \frac{2\sqrt{2}\rho^2}{\overline{\rho}\Delta}\Bigg[\left(\frac{i\mathcal{K}}{\Delta}+\rho^{-1}+\overline{\rho}^{-1}\right)L_2 \\
&\hspace{1.5cm}- a\sqrt{1-z^2}\frac{\mathcal{K}}{\Delta}\left(\overline{\rho}^{-1} - \rho^{-1}\right)\Bigg]\frac{S^{aw}_{lm}}{\sqrt{2\pi}}, \\
f^{(0)}_{\overline{m}\overline{m}} &= \frac{\rho^2}{\overline{\rho}^2}\left[i\partial_r\frac{\mathcal{K}}{\Delta} - 2i\rho^{-1}\frac{\mathcal{K}}{\Delta} + \frac{\mathcal{K}^2}{\Delta^2}\right]\frac{S^{aw}_{lm}}{\sqrt{2\pi}}, \\
f^{(1)}_{n\overline{m}} &= \frac{2\sqrt{2}\rho^2}{\overline{\rho}\Delta}\left[L_2 + ia\sqrt{1-z^2}\left(\overline{\rho}^{-1} - \rho^{-1}\right)\right]\frac{S^{aw}_{lm}}{\sqrt{2\pi}}, \\
f^{(1)}_{\overline{m}\overline{m}} &= -\frac{2\rho^2}{\overline{\rho}^2}\left(\rho^{-1} + i\frac{\mathcal{K}}{\Delta}\right)\frac{S^{aw}_{lm}}{\sqrt{2\pi}}, \\
f^{(2)}_{\overline{m}\overline{m}} &= -\frac{\rho^2}{\overline{\rho}^2}\frac{S^{aw}_{lm}}{\sqrt{2\pi}},
\end{aligned}
\end{equation}
where $L_n := \partial_{\theta} - m/\sin\theta + aw\sin\theta + n\cot\theta$.

Via the Green function method, the solution of Eq.~\ref{eq:RadialEq} can be constructed by two linearly independent homogeneous solutions, $R^{\text{in}}_{lmw}$ and $R^{\text{up}}_{lmw}$
\begin{equation}
\begin{aligned}
R_{lmw}(r) =& \frac{R^{\text{up}}_{lmw}(r)}{W}\int_{r_{+}}^{r}dr' \ \frac{T_{lmw}(r')R^{\text{in}}_{lmw}(r')}{\Delta^2(r')} \\
&+ \frac{R^{\text{in}}_{lmw}(r)}{W}\int_{r}^{\infty}dr' \ \frac{T_{lmw}(r')R^{\text{up}}_{lmw}(r')}{\Delta^2(r')},
\end{aligned}
\end{equation}
where $W$ is the Wronskian constant, and the boundary condition of homogeneous solutions is set as
\begin{equation}
R^\text{in}_{lmw} \rightarrow 
\begin{cases}
\begin{aligned}
B^\text{trans}\Delta^2 e^{-ik_wr_*} \ \ \ &\text{for} \ r\rightarrow r_+ \\
r^3 B^\text{out} e^{iwr_*} + r^{-1} B^\text{in} e^{-iwr_*} \ \ \ &\text{for} \ r \rightarrow \infty
\end{aligned},
\end{cases}
\end{equation}
\begin{equation}
R^\text{up}_{lmw} \rightarrow 
\begin{cases}
\begin{aligned}
C^\text{out}e^{ik_wr_*} + \Delta^2 C^\text{in} e^{-ik_wr_*} \ \ \ &\text{for} \ r\rightarrow r_+ \\
C^\text{trans}r^3 e^{iwr_*} \ \ \ &\text{for} \ r \rightarrow \infty
\end{aligned},
\end{cases}
\end{equation}
where $k_w:=w-ma/(2Mr_+)$ and $r_*$ is the torroise coordinate $dr_*/dr \equiv (r^2+a^2)/\Delta$. Thus, the asymptotic property of the radial solution at the horizon and infinity is
\begin{equation}\label{eq:amplitudeREQH}
\begin{aligned}
R_{lmw}(r \rightarrow r_+) &= \frac{B^\text{trans}_{lmw}\Delta^2 e^{-ikr_*}}{2iwC^\text{trans}_{lmw}B^\text{in}_{lmw}} \int_{r_+}^{\infty} dr' \frac{R^\text{up}_{lmw}T_{lmw}}{\Delta^2} \\
&\equiv Z^{\text{H}}_{lmw}\Delta^2 e^{-ikr_*},
\end{aligned} 
\end{equation}
\begin{equation}\label{eq:amplitudeREQ8}
\begin{aligned}
R_{lmw}(r \rightarrow \infty) &= \frac{r^3 e^{iwr_*}}{2iwB^\text{in}_{lmw}} \int_{r_+}^{\infty} dr' \frac{R^\text{in}_{lmw}T_{lmw}}{\Delta^2} \\
&\equiv Z^{\infty}_{lmw}r^3 e^{iwr_*}.
\end{aligned} 
\end{equation}

In this paper, the numerical solutions of polar functions $S^{aw}_{lm}(z)$ and homogeneous radial functions $R^{\text{in/up}}_{lmw}(r)$ are calculated through the Jiang-Han method \cite{Jiang2026} by their RUST codes. Therefore, to compute the asymptotic gravitational perturbation, the only remaining step that has not been completed is submitting the corresponding source term $T_{lmw}(r)$ into the integration above to get the amplitude $Z^{{\infty/\text{H}}}_{lmw}$.

\subsection{Source term}
Firstly, let us consider the general expression of $T^{\mu\nu}$ as
\begin{equation}
\begin{aligned}
T^{\mu\nu}(x^{\mu}) = &\int d\tau\Bigg[\frac{\delta^{(4)}(x^{\mu}-z^{\mu}(\tau))}{\sqrt{-g}}\mathfrak{t}^{\mu\nu}_0(z^{\mu}(\tau))\\
&-\nabla_{\rho}\left(\frac{\delta^{(4)}(x^{\mu}-z^{\mu}(\tau))}{\sqrt{-g}}\mathfrak{t}^{\mu\nu\rho}_1(z^{\mu}(\tau))\right) \\
&+\nabla_{\rho}\nabla_{\sigma}\left(\frac{\delta^{(4)}(x^{\mu}-z^{\mu}(\tau))}{\sqrt{-g}}\mathfrak{t}^{\mu\nu\rho\sigma}_2(z^{\mu}(\tau))\right)\Bigg].
\end{aligned}
\end{equation}
Project it onto the null tetrad and use $\Gamma^{\mu}_{\mu\alpha}\equiv\frac{1}{\sqrt{-g}}\partial_{\alpha}\sqrt{-g}$
\begin{equation}
\begin{aligned}
T_{ab} &\equiv T_{\mu\nu}\lambda^{\mu}_a\lambda^{\nu}_b \\
&= \frac{1}{\sqrt{-g}}\int d\tau \ \bigg[A_{ab}\delta^{(4)}(x^{\mu}-z^{\mu}(\tau)) \\
&\hspace{1cm}-\partial_{\rho}\left(B^{\rho}_{ab}\delta^{(4)}(x^{\mu}-z^{\mu}(\tau))\right) \\
&\hspace{1cm}+\partial_{\rho}\partial_{\sigma}\left(C^{\rho\sigma}_{ab}\delta^{(4)}(x^{\mu}-z^{\mu}(\tau))\right)\bigg],
\end{aligned}
\end{equation}
where
\begin{equation}
\begin{aligned}
A_{ab} &= \mathfrak{t}_{0}^{\mu\nu}\lambda_{a\mu}\lambda_{b\nu} + \mathfrak{t}^{\mu\nu\rho}_{1}\nabla_{\rho}\left(\lambda_{a\mu}\lambda_{b\nu}\right) + \mathfrak{t}^{\mu\nu\rho\sigma}_2\nabla_{\sigma}\nabla_{\rho}\left(\lambda_{a\mu}\lambda_{b\nu}\right), \\
B^{\rho}_{ab} &= \mathfrak{t}_{1}^{\mu\nu\rho}\lambda_{a\mu}\lambda_{b\nu} + 2\mathfrak{t}_2^{\mu\nu(\rho\sigma)}\nabla_{\sigma}\left(\lambda_{a\mu}\lambda_{b\nu}\right) - \Gamma^{\rho}_{\gamma\sigma}\mathfrak{t}_{2}^{\mu\nu\gamma\sigma}\lambda_{a\mu}\lambda_{b\nu},\\
C^{\rho\sigma}_{ab} &= \mathfrak{t}_{2}^{\mu\nu\rho\sigma}\lambda_{a\mu}\lambda_{b\nu}.
\end{aligned}
\end{equation}
As mentioned previously, the stress-energy tensor of a spinning Dixon particle is approximately in the following form:
\begin{equation}\label{eq:stressEnergyT}
\begin{aligned}
T^{\mu\nu}(x^{\mu}) = &m_s \int d\tau \ \Bigg[\frac{\delta^{(4)}(x^{\mu}-z^{\mu}(\tau))}{\sqrt{-g}}v^{\mu}v^{\nu} \\
&- \nabla_{\rho}\left(\frac{\delta^{(4)}(x^{\mu}-z^{\mu}(\tau))}{\sqrt{-g}}s^{\rho(\mu}v^{\nu )}\right)\Bigg].
\end{aligned}
\end{equation}
It corresponds to the case where $\mathfrak{t}^{\mu\nu}_0=m_sv^{\mu}v^{\nu}$, $\mathfrak{t}^{\mu\nu\rho}_1=m_ss^{\rho(\mu}v^{\nu )}$ and $\mathfrak{t}^{\mu\nu\rho\sigma}_2=0$.

Submitting the source term into the expression of amplitude in Eq.~\ref{eq:amplitudeREQH} and Eq.~\ref{eq:amplitudeREQ8}, we have
\begin{equation}\label{eq:SourceTermIntegration}
\begin{aligned}
Z^{\text{H}/\infty}_{lmw} =& C_{\text{H}/\infty}\int_{-\infty}^{\infty} d\tau \ e^{iwt(\tau)-im\varphi(\tau)}\mathcal{I}^{\text{H}/\infty}_{lmw}(r(\tau),z(\tau),v^{\alpha}(\tau),s^{\alpha}(\tau)), \\
\mathcal{I}^{\text{H}/\infty}_{lmw} :=&\sum_{ab}\sum_{i=0}^{N_{ab}}(-1)^i \Bigg[ \left(A_{ab}f^{(i)}_{ab} + B^{\rho}_{ab}\partial_{\rho}f^{(i)}_{ab} + C^{\rho\sigma}_{ab}\partial_{\sigma}\partial_{\rho}f^{(i)}_{ab} \right)\frac{d^i R^{\text{up/in}}_{lmw}}{dr^i} \\
\ \ \ \ \ \ \ \ \ \ \ &+ \left(B^r_{ab}f^{(i)}_{ab} + C^{(r\rho)}_{ab}\partial_{\rho}f^{(i)}_{ab} \right)\frac{d^{i+1}R^{\text{up/in}}_{lmw}}{dr^{i+1}} + C^{rr}_{ab}f^{(i)}_{ab}\frac{d^{i+2}R^{\text{up/in}}_{lmw}}{dr^{i+2}} \Bigg],
\end{aligned}
\end{equation}
where $C_{\text{H}}:=B^{\text{trans}}_{lmw}/(2iwC^{\text{trans}}_{lmw}B^{\text{in}}_{lmw})$, $C_{\infty}:=1/(2iwB^{\text{in}}_{lmw})$. To reach the equation above, the method of integration by parts is used. It should be noted that, here we use $\partial_tf^{(i)}_{ab}$ to label $iwf^{(i)}_{ab}$ and $\partial_{\varphi}f^{(i)}_{ab}$ to label $-imf^{(i)}_{ab}$ for the sake of simplicity although the true derivative must be equal to 0, and the $iw$ and $-im$ actually come from the derivative of $e^{iwt-im\varphi}$, which should not cause any ambiguity.

To numerically calculate the integration from infinity to infinity in Eq.~\ref{eq:SourceTermIntegration}, we need to apply the Fourier expansion. Using the MPD orbit and motion, this integration can be reorganized in the following form:
\begin{equation}\label{eq:SourceTermIntegration2}
Z^{\text{H}/\infty}_{lmw} = C_{\text{H}/\infty}\int^{\infty}_{-\infty}d\tilde{\lambda}e^{i(w\tilde{\Gamma}-m\tilde{\Upsilon}_{\varphi})\tilde{\lambda}}\mathcal{J}^{\text{H}/\infty}_{lmw}\left(\tilde{r}(\tilde{\lambda}),\tilde{z}(\tilde{\lambda}),\psi(\tilde{\lambda})\right),
\end{equation}
where
\begin{equation}
\mathcal{J}^{\text{H}/\infty}_{lmw} := \frac{d\tau}{d\tilde{\lambda}}e^{iw\left[\tilde{t}^{(r)}+\tilde{t}^{(z)}-\delta t(\tilde{r},\tilde{z},\psi)\right]-im\left[\tilde{\varphi}^{(r)}+\tilde{\varphi}^{(z)}-\delta\varphi(\tilde{r},\tilde{z},\psi)\right]}\mathcal{I}^{\text{H}/\infty}_{lmw}(\tilde{r},\tilde{z},\psi).
\end{equation}
Then we perform 3D Fourier expansion to $\mathcal{J}^{\text{H}/\infty}_{lmw}$ with $\{q_{\tilde{r}}:=\tilde{\Upsilon}_r\tilde{\lambda},q_{\tilde{z}}:=\tilde{\Upsilon}_z\tilde{\lambda},q_{\psi}:=\Upsilon_{\psi}\tilde{\lambda}\}$
\begin{equation}\label{eq:3DFourierExpansion}
\mathcal{J}^{\text{H}/\infty}_{lmknj} = \frac{1}{(2\pi)^3}\int_{-\pi}^{\pi}dq_{\tilde{z}}\int_{-\pi}^{\pi}dq_{\tilde{z}}\int_{-\pi}^{\pi}dq_{\psi} \ \mathcal{J}^{\text{H}/\infty}_{lmw}e^{i(k\tilde{\Upsilon}_z+n\tilde{\Upsilon}_r+j\Upsilon_{\psi})\tilde{\lambda}}.
\end{equation}
Thus, we have
\begin{equation}\label{eq:FourierEXP}
\mathcal{J}^{\text{H}/\infty}_{lmw} = \sum_{k,n=-\infty}^{\infty}\sum_{j=-1}^{1}\mathcal{J}^{\text{H}/\infty}_{lmknj}e^{-i(k\tilde{\Upsilon}_z+n\tilde{\Upsilon}_r+j\Upsilon_{\psi})\tilde{\lambda}},
\end{equation}
where $j \in \{-1,0,1\}$ since $q_{\psi}$ only appears in the form of $\sin$ and $\cos$. And if $s_{\perp}=0$, $\psi$ disappears in the source term, then only $\mathcal{J}^{\text{H}/\infty}_{lmkn0}$ is nonzero. Submitting Eq.~\ref{eq:FourierEXP} into Eq.~\ref{eq:SourceTermIntegration2} and using the Fourier property of $\delta(x)$ function, we have
\begin{equation}
Z^{\text{H}/\infty}_{lmw} = C_{\text{H}/\infty}\sum_{k,n=-\infty}^{\infty}\sum_{j=-1}^{1}\frac{2\pi}{\tilde{\Gamma}}\mathcal{J}^{\text{H}/\infty}_{lmknj} \ \delta(w-w_{mknj}),
\end{equation}
where $w_{mknj}:=(m\tilde{\Upsilon}_{\varphi}+k\tilde{\Upsilon}_z+n\tilde{\Upsilon}_r+j\Upsilon_{\psi})/\tilde{\Gamma}$.

\subsection{Asymptotic wave and radiative field}
The asymptotic wave can be recovered from $\Psi_4$ as
\begin{equation}
\Psi_4(r\rightarrow\infty) = \frac{1}{2}\frac{\partial^2}{\partial t^2}\left(h_+ - ih_{\times}\right).
\end{equation}
Since $Z^{\infty}_{lmw}$ is composed of discrete spectra with
\begin{equation}\label{eq:discreteAmp}
\begin{aligned}
Z^{\infty}_{lmw} &= \sum_{k,n=-\infty}^{\infty}\sum_{j=-1}^{1}Z^{\infty}_{lmknj}\delta(w-w_{mknj}), \\
Z^{\infty}_{lmknj} &= \frac{2\pi C_{\infty}}{\tilde{\Gamma}}\mathcal{J}^{\infty}_{lmknj},
\end{aligned}
\end{equation}
the $\Psi_4(r\rightarrow\infty)$ dose so with
\begin{equation}
\begin{aligned}
\Psi_4(r\rightarrow\infty) &= \sum_{l=2}^{\infty}\sum_{m=-l}^{l}\sum_{k,n=-\infty}^{\infty}\sum_{j=-1}^{1} \Psi_{4,lmknj}, \\
\Psi_{4,lmknj}(r\rightarrow\infty) &= \lim_{r\rightarrow\infty}\frac{r^3}{\rho^4} Z^{\infty}_{lmknj}e^{-iw_{mknj}(t-r_*)+im\varphi}\frac{S^{aw_{mknj}}_{lm}}{\sqrt{2\pi}}.
\end{aligned}
\end{equation}
Thus, the polarization of the gravitational wave can be calculated from
\begin{equation}
h_+ - ih_{\times} = \sum_{l=2}^{\infty}\sum_{m=-l}^{l}\sum_{k,n=-\infty}^{\infty}\sum_{j=-1}^{1}\frac{-2}{w^2_{lmknj}} \Psi_{4,lmknj}.
\end{equation}

In Fig.~\ref{fig:snapshot}, we show an example of snapshot waveforms with a spinning secondary. $\{E,J_z,K\}$ are set to share the same value with a geodesic of $\{a=0.9M,p=6M,e=0.1,\theta_\text{min}=70^{\circ}\}$. The field point at which the gravitational perturbation is evaluated is set to $\{r\rightarrow\infty,z=1/2,\varphi=\pi/2\}$. The Fourier expansion in Eq.~\ref{eq:3DFourierExpansion} is realized by the Trapezoidal Rule with a fixed grid of $100 \times 100$ uniformly spaced points. The truncation of the mode number is fixed at $\left\{l_{\text{max}} = 20, k_{\text{max}} = 10, n_{\text{max}}=10 \right\}$. Those settings will be used repeatedly in the following sections.
\begin{figure*}
    \centering
    \includegraphics[width=1.\linewidth]{waveform_snapshot.jpg}
    \caption{The snapshot EMRI waveform with a spinning secondary. The orbital parameters are fixed as $a=0.9M$, $s_{||}=0.00001M$, $s_{\perp}=0$, $E=9.5286602596\times 10^{-1}$, $J_z=3.2638133216M$ and $K=7.2098629692M^2$.}
    \label{fig:snapshot}
\end{figure*}

In addition to the physical metric perturbation, the radiative metric perturbation (i.e., $h^{\text{rad}}_{\mu\nu}$) can also be reconstructed from the radiative solution of Teukolsky equations, which is defined as the retarded field minus the
advanced field \cite{Chrzanowski1975,Wald1978,Galtsov1982}. The derivation is very well summarized in Ref.~\cite{Sago2006}.
\begin{equation}\label{eq:RadFieldMetric}
\begin{aligned}
h^{\text{rad}}_{\mu\nu}(x) = \int dw &\sum_{lm}\frac{1}{2iw^3}\Bigg(N_{\text{out}}\Pi^{\text{out}}_{lmw,\mu\nu}(x)\\
&\int d^4x'\sqrt{-g} \ \overline{N}_{\text{out}}\overline{\Pi}^{\text{out}}_{lmw,\alpha\beta}(x')T^{\alpha\beta}(x') \\
&+\frac{w}{k_w}N_{\text{down}}\Pi^{\text{down}}_{lmw,\mu\nu}(x)\\
&\int d^4x'\sqrt{-g} \ \overline{N}_{\text{down}}\overline{\Pi}^{\text{down}}_{lmw,\alpha\beta}(x')T^{\alpha\beta}(x')\Bigg) + (c.c.),
\end{aligned}
\end{equation}
where $(c.c.)$ refers to the complex conjugation part which keeps the metric solution real, and
\begin{equation}\label{eq:PiOUTandDOWN}
\Pi^{\text{out/down}}_{lmw,\mu\nu} := \frac{16}{\overline{\mathcal{C}}} \tau^*_{\mu\nu}\left[\frac{\overline{R}^{\text{in/up}}_{lmw}}{\Delta^2}\frac{S^{aw}_{lm}}{\sqrt{2\pi}}e^{im\varphi-iwt}\right],
\end{equation}
with the constant $N_{\text{out/down}}$ and $\mathcal{C}$ 
\begin{equation}\label{eq:NoutNdown}
\begin{aligned}
|N_{\text{out}}|^2 =& \frac{|\mathcal{C}|^2}{256w^2|B^{\text{in}}_{lmw}|^2}, \\
|N_{\text{down}}|^2 =& \frac{k^6_w|\kappa_1|^2|\kappa_2|^2(2Mr_+)^5|B^{\text{trans}}_{lmw}|^2}{|B^{\text{in}}_{lmw}|^2 |C^{\text{trans}}_{lmw}|^2}, \\
\kappa_s :=& 1 - \frac{is(r_+-M)}{2k_wMr_+}, \\
\mathcal{C} =& 12iwM +\bigg[\left(\left(\lambda_{lmw}+2\right)^2+4awm-4a^2w^2\right)\\
&\ \ \ \ \ \ \left(\lambda_{lmw}^2+36awm-36a^2w^2\right) \\
&+\left(2\lambda_{lmw}+3\right)\left(96a^2w^2-48awm\right) - 144a^2w^2\bigg]^{1/2}.
\end{aligned}
\end{equation}
$\tau^*_{\mu\nu}$ is the adjoint of the operator $\tau_{\mu\nu}$, defined as satisfying
\begin{equation}
\int \sqrt{-g} \ \overline{X}\tau_{\mu\nu}Y^{\mu\nu}d^4x = \int \sqrt{-g}Y^{\mu\nu}\overline{\tau^*_{\mu\nu}X}, 
\end{equation}
for an arbitrary scalar field $X$ and tensor field $Y^{\mu\nu}$.
Using this property, the inner integration in Eq.~\ref{eq:RadFieldMetric} can be simplified in form of the amplitude that we introduce in Eq.~\ref{eq:amplitudeREQH} and Eq.~\ref{eq:amplitudeREQ8}
\begin{equation}\label{eq:amplitudeRad}
\begin{aligned}
&\int d^4x'\sqrt{-g} \ \overline{N}_{\text{out/down}}\overline{\Pi}^{\text{out/down}}_{lmw,\alpha\beta}T^{\alpha\beta} \\
=& \frac{16\overline{N}_{\text{out/down}}}{\mathcal{C}}\int d^4x'\sqrt{-g}\frac{R^{\text{in/up}}_{lmw}}{\Delta^2}\frac{\overline{S}^{aw}_{lm}}{\sqrt{2\pi}}e^{iwt-im\varphi}\tau_{\mu\nu}\left[T^{\mu\nu}\right] \\
=& \frac{16\overline{N}_{\text{out/down}}}{\mathcal{C}}\int d^4x' \frac{R^{\text{in/up}}_{lmw}}{\Delta^2}\frac{\overline{S}^{aw}_{lm}}{\sqrt{2\pi}}e^{iwt-im\varphi} \\
&\ \ \ \ \frac{1}{4\pi}\int dw'\sum_{l'm'}T_{l'm'w'}e^{im'\varphi-iw't}\frac{S^{aw}_{lm}}{\sqrt{2\pi}} \\
=& \frac{8\overline{N}_{\text{out/down}}}{C_{\infty/\text{H}}\mathcal{C}}Z^{\infty/\text{H}}_{lmw}.
\end{aligned}
\end{equation}
Specifically, for the out case
\begin{equation}
\vert\frac{8\overline{N}_{\text{out}}}{C_{\infty}\mathcal{C}}\vert^2 = 1,
\end{equation}
for the down case
\begin{equation}
\vert\frac{8\overline{N}_{\text{down}}}{C_{\text{H}}\mathcal{C}}\vert^2 = \frac{256w^2k^6_w\vert\kappa_1\vert^2\vert\kappa_2\vert^2\left(2Mr_+\right)^5}{\vert\mathcal{C}\vert^2}.
\end{equation}
Commonly, we define $\alpha_{lmw} :=\frac{w}{k_w}\vert\frac{8\overline{N}_{\text{down}}}{C_{\text{H}}\mathcal{C}}\vert^2$.

\section{Fluxes of Gravitational Radiation}\label{Sec:OrbitInspiral}
In self-force theory, the equations of motion for the momentum $P^{\mu}$ and the spin tensor $S^{\mu\nu}$ are still described by Eq.~\ref{eq:MPDEQ} but in an effective metric $\hat{g}_{\alpha\beta} = g_{\alpha\beta} + h^E_{\alpha\beta}$ instead of the background (Kerr) metric $g_{\alpha\beta}$ \cite{Harte2012,Mathews2022,Barack2019},
\begin{equation}
\begin{aligned}
\frac{\hat{D}P^{\mu}}{d\hat{\tau}} &= -\frac{1}{2}\hat{R}^{\mu}_{\ \ \nu\rho\sigma}\hat{v}^{\nu}S^{\rho\sigma}, \\
\frac{\hat{D}S^{\mu\nu}}{d\hat{\tau}} &= 2P^{[\mu}\hat{v}^{\nu]}.
\end{aligned}
\end{equation}
We again adopt the TD condition and apply the linear-spin approximation
\begin{equation}
\begin{aligned}
\frac{\hat{D}\hat{v}^{\mu}}{d\hat{\tau}} &= -\frac{1}{2}\hat{R}^{\mu}_{\ \ \nu\rho\sigma}\hat{v}^{\nu}s^{\rho\sigma}, \\
\frac{\hat{D}s^{\mu\nu}}{d\hat{\tau}} &= 0,
\end{aligned}
\end{equation}
where $s^{\rho\sigma} = S^{\rho\sigma}/\hat{m}_s$. Expanding the above equations in powers of $h^E_{\alpha\beta}$ and only keeping the linear term, we will get
\begin{equation}
\begin{aligned}
f^{\mu}=& -\frac{1}{2}P^{\mu\lambda}\left(2h^{E}_{\lambda\rho ; \sigma}-h^{E}_{\rho\sigma ; \lambda}\right)v^{\rho}v^{\sigma} \\
&-\frac{1}{2}R^{\mu}_{\ \ \alpha\beta\gamma}\left(1-\frac{1}{2}h^{E}_{\rho\sigma}u^{\rho}u^{\sigma}\right)v^{\alpha}s^{\beta\gamma} \\
&+\frac{1}{2}P^{\mu\nu}\left(2h^{E}_{\nu(\alpha ; \beta)\gamma}-h^{E}_{\alpha\beta ; \nu\gamma}\right)v^{\alpha}s^{\beta\gamma}, \\
n^{\mu\nu} =& v^{(\rho}s^{\sigma )[\mu}g^{\nu ]\lambda}\left(2h^{E}_{\lambda\rho ; \sigma}-h^{E}_{\rho\sigma ; \lambda}\right), \\
P^{\mu\nu} :=& g^{\mu\nu} + v^{\mu}v^{\nu},
\end{aligned}
\end{equation}
where we define the self-force $f^{\mu}:=Dv^{\mu}/d\tau$ and the self-torque $n^{\mu\nu} := Ds^{\mu\nu}/d\tau$. We reorganize the self-force and self-torque according to the different derivative orders of the metric perturbation as
\begin{equation}\label{eq:self-force-form}
\begin{aligned}
f^{\mu} &= -\frac{1}{2}R^{\mu}_{\alpha\beta\gamma}u^{\alpha}s^{\beta\gamma} + \mathfrak{t}^{\mu\alpha\beta}_{0,f}h^E_{\alpha\beta} + \mathfrak{t}^{\mu\alpha\beta\gamma}_{1,f}h^E_{\alpha\beta;\gamma} + \mathfrak{t}^{\mu\alpha\beta\gamma\sigma}_{2,f}h^E_{\alpha\beta;\gamma\sigma}, \\
n^{\mu\nu} &= \mathfrak{t}^{\mu\nu\alpha\beta\gamma}_{1,n}h^E_{\alpha\beta;\gamma}.
\end{aligned}
\end{equation}

In order to evolve the MPD orbit, we need to calculate the the orbit-averaged rates of change of four constants of motion, i.e., $C = \left\{E,J_z,K,s_{||}\right\}$
\begin{equation}
\left<\frac{dC}{d\tilde{\lambda}}\right> = \lim_{T\rightarrow\infty}\frac{1}{2T}\int_{-T}^Td\tilde{\lambda}\frac{d\tau}{d\tilde{\lambda}}\left(\frac{\partial C}{\partial v^{\alpha}}f^{\alpha} + \frac{\partial C}{\partial s^{\alpha\beta}}n^{\alpha\beta}\right),
\end{equation}
where we use $<\dots>$ to label the orbit-averaged operation $\lim_{T\rightarrow\infty}\frac{1}{2T}\int_{-T}^{T}d\tilde{\lambda}\frac{d\tau}{\tilde{d\lambda}}\left(\dots\right)$.
Substituting Eq.~\ref{eq:self-force-form}, we obtain
\begin{equation}\label{eq:ConstantEvolve}
\begin{aligned}
\left<\frac{dC}{d\tilde{\lambda}}\right> =& \Bigg<\frac{\partial C}{\partial v^{\mu}}\mathfrak{t}^{\mu\alpha\beta}_{0,f}h^E_{\alpha\beta} + \left(\frac{\partial C}{\partial v^{\mu}}\mathfrak{t}^{\mu\alpha\beta\gamma}_{1,f} + \frac{\partial C}{\partial s^{\mu\nu}}\mathfrak{t}^{\mu\nu\alpha\beta\gamma}_{1,n}\right)h^E_{\alpha\beta;\gamma}\\
&\frac{\partial C}{\partial v^{\mu}}\mathfrak{t}^{\mu\alpha\beta\gamma\sigma}_{2,f}h^E_{\alpha\beta;\gamma\sigma}\Bigg>,
\end{aligned}
\end{equation}
where $-\frac{1}{2}R^{\mu}_{\alpha\beta\gamma}v^{\alpha}s^{\beta\gamma}$ corresponds to the background terms, thus it is naturally eliminated. As proposed by Mino \cite{Mino2003,VanDeMeent2018}, the radiative field $h^{\text{rad}}_{\mu\nu}$ can be used to calculate the evolution in the orbit-average sense, i.e., $h^E_{\mu\nu} \approx h^{\text{rad}}_{\mu\nu}$. When considering the radiative field in Eq.~\ref{eq:RadFieldMetric}, there exists a direct way to numerically calculate Eq.~\ref{eq:ConstantEvolve}.
We rewrite it in the following form:
\begin{equation}
\left<\frac{dC}{d\tilde{\lambda}}\right> = \left<h^{\text{rad}}_{\mu\nu}\mathfrak{t}^{\mu\nu}_0 + h^{\text{rad}}_{\mu\nu;\rho}\mathfrak{t}^{\mu\nu\rho}_1 + h^{\text{rad}}_{\mu\nu;\rho\sigma}\mathfrak{t}^{\mu\nu\rho\sigma}_2\right>.
\end{equation}
For the first term about $\mathfrak{t}^{\mu\nu}_0$
\begin{equation}
\begin{aligned}
\left<h^{\text{rad}}_{\mu\nu}\mathfrak{t}^{\mu\nu}_0\right> = \bigg<\int dw &\sum_{lm} \frac{1}{2iw^3}\bigg(N_{\text{out}}\Pi^{\text{out}}_{lmw,\mu\nu}(z^{\mu}(\tilde{\lambda}))\mathfrak{t}^{\mu\nu}_0(z^{\mu}(\tilde{\lambda}))Z^{\infty}_{lmw} \\
& + \text{the "down" part}\bigg) + (c.c.)\bigg>.
\end{aligned}
\end{equation}
We define the "out" part of the above equation as
\begin{equation}
\phi^{\text{out}}_{lmw}(\mathfrak{t}^{\mu\nu}_0,\tilde{\lambda}) := \frac{d\tau}{d\tilde{\lambda}}N_{\text{out}}\Pi^{\text{out}}_{lmw,\mu\nu}(z^{\mu}(\tilde{\lambda}))\mathfrak{t}^{\mu\nu}_0(z^{\mu}(\tilde{\lambda})).
\end{equation}
Performing the Fourier transformation to it
\begin{equation}
\begin{aligned}
\phi^{\text{out}}_{lmw}(\mathfrak{t}^{\mu\nu}_0,\tilde{\lambda}) =& \frac{1}{2\pi}\int d\eta \Phi^{\text{out}}_{lmw}(\eta)e^{i\eta\tilde{\lambda}}, \\
\Phi^{\text{out}}_{lmw}(\mathfrak{t}^{\mu\nu}_0 ,\eta) :=& \int d\tilde{\lambda} \phi^{\text{out}}_{lmw}(\mathfrak{t}^{\mu\nu}_0,\tilde{\lambda})e^{-i\eta\tilde{\lambda}}.
\end{aligned}
\end{equation}
Using the Dirac function $\delta^{(4)}$, $\Phi^{\text{out}}_{lmw}$ can be reorganized into a form similar to Eq.~\ref{eq:amplitudeRad}
\begin{equation}\label{eq:Phiout}
\begin{aligned}
\Phi^{\text{out}}_{lmw}(\mathfrak{t}^{\mu\nu}_0,\eta) =  \int d^4x &\sqrt{-g} N_{\text{out}}\Pi^{\text{out}}_{lmw,\mu\nu}(x^{\mu}) \\
&\int \frac{d\tau}{d\tilde{\lambda}}d\tilde{\lambda}\frac{\delta^{(4)}(x^{\mu}-z^{\mu}(\tilde{\lambda}))}{\sqrt{-g}}\mathfrak{t}^{\mu\nu}_0(z^{\mu}(\tilde{\lambda}))e^{-i\eta\tilde{\lambda}}.
\end{aligned}
\end{equation}
If we treat $\int d\tau \delta^{(4)}\mathfrak{t}^{\mu\nu}_0/\sqrt{-g}$ as the virtual stress-energy tensor and label the corresponding virtual amplitude (in Eq.~\ref{eq:discreteAmp}) as $Z^{\infty/\text{H},C_0}_{lmknj}$, we have
\begin{equation}
\Phi^{\text{out}}_{lmw}(\mathfrak{t}_0^{\mu\nu},\eta) = \frac{16\overline{N}_{\text{out}}}{C_{\infty}\mathcal{C}}\tilde{\Gamma}\sum_{knj}\overline{Z}^{\infty,C_0}_{lmknj}\delta(w\tilde{\Gamma}-m\tilde{\Upsilon}_{\varphi}-k\tilde{\Upsilon}_z-n\tilde{\Upsilon}_r-j\tilde{\Upsilon}_{\psi}-\eta),
\end{equation}
where the $-\eta$ in the $\delta$ function is induced by $e^{-i\eta\tilde{\lambda}}$ in Eq.~\ref{eq:Phiout}. So
\begin{equation}
\phi^{\text{out}}_{lmw}(\mathfrak{t}^{\mu\nu}_0,\tilde{\lambda}) = \frac{\tilde{\Gamma}}{2\pi}\sum_{knj}\overline{Z}^{\infty,C_0}_{lmknj}e^{i\tilde{\Gamma}(w-w_{mknj})\tilde{\lambda}}.
\end{equation}
For terms about $\mathfrak{t}^{\mu\nu\rho}_1$ or $\mathfrak{t}^{\mu\nu\rho\sigma}_2$, the situation is quite similar by using the Stokes' theorem
\begin{equation}
\begin{aligned}
\Phi^{\text{out}}_{lmw}(\mathfrak{t}^{\mu\nu\rho}_1,\eta) &= \int \frac{d\tau}{d\tilde{\lambda}} N_{\text{out}}d\tilde{\lambda}\int d^4x\sqrt{-g}\Pi^{\text{out}}_{lmw,\mu\nu;\rho}(x^{\mu}) \\
&\ \ \ \ \ \ \ \frac{\delta^{(4)}(x^{\mu}-z^{\mu}(\tilde{\lambda}))}{\sqrt{-g}}\mathfrak{t}^{\mu\nu\rho}_1(z^{\mu}(\tilde{\lambda}))e^{-i\eta\tilde{\lambda}} \\
&= \int d^4x\sqrt{-g}N_{\text{out}}\Pi^{\text{out}}_{lmw}(x^{\mu}) \\
&\ \ \ \ \ \ \ \int \frac{d\tau}{d\tilde{\lambda}} d\tilde{\lambda}\nabla_{\rho}\left(\frac{-\delta^{(4)}(x^{\mu}-z^{\mu}(\tilde{\lambda}))}{\sqrt{-g}}\mathfrak{t}^{\mu\nu\rho}_1(z^{\mu}(\tilde{\lambda}))e^{-i\eta\tilde{\lambda}}\right),
\end{aligned}
\end{equation}
\begin{equation}
\begin{aligned}
\Phi^{\text{out}}_{lmw}(\mathfrak{t}^{\mu\nu\rho\sigma}_2,\eta) &= \int \frac{d\tau}{d\tilde{\lambda}} N_{\text{out}}d\tilde{\lambda}\int d^4x\sqrt{-g}\Pi^{\text{out}}_{lmw,\mu\nu;\rho\sigma}(x^{\mu}) \\
&\ \ \ \ \ \ \ \frac{\delta^{(4)}(x^{\mu}-z^{\mu}(\tilde{\lambda}))}{\sqrt{-g}}\mathfrak{t}^{\mu\nu\rho\sigma}_2(z^{\mu}(\tilde{\lambda}))e^{-i\eta\tilde{\lambda}} \\
&= \int d^4x\sqrt{-g}N_{\text{out}}\Pi^{\text{out}}_{lmw}(x^{\mu}) \\
&\int \frac{d\tau}{d\tilde{\lambda}} d\tilde{\lambda}\nabla_{\sigma}\nabla_{\rho}\left(\frac{\delta^{(4)}(x^{\mu}-z^{\mu}(\tilde{\lambda}))}{\sqrt{-g}}\mathfrak{t}^{\mu\nu\rho\sigma}_2(z^{\mu}(\tilde{\lambda}))e^{-i\eta\tilde{\lambda}}\right).
\end{aligned}
\end{equation}
Therefore, for the full terms $\left<h^{\text{rad}}_{\mu\nu}\mathfrak{t}^{\mu\nu}_0 + h^{\text{rad}}_{\mu\nu;\rho}\mathfrak{t}^{\mu\nu\rho}_1 + h^{\text{rad}}_{\mu\nu;\rho\sigma}\mathfrak{t}^{\mu\nu\rho\sigma}_2\right>$, the virtual stress-energy tensor is
\begin{equation}\label{eq:virtualTensor}
\begin{aligned}
T^{\mu\nu}_{\text{C}}(x^{\mu}) = &\int d\tau\Bigg[\frac{\delta^{(4)}(x^{\mu}-z^{\mu}(\tau))}{\sqrt{-g}}\mathfrak{t}^{\mu\nu}_0(z^{\mu}(\tau)) \\
&-\nabla_{\rho}\left(\frac{\delta^{(4)}(x^{\mu}-z^{\mu}(\tau))}{\sqrt{-g}}\mathfrak{t}^{\mu\nu\rho}_1(z^{\mu}(\tau))\right) \\
&+\nabla_{\rho}\nabla_{\sigma}\left(\frac{\delta^{(4)}(x^{\mu}-z^{\mu}(\tau))}{\sqrt{-g}}\mathfrak{t}^{\mu\nu\rho\sigma}_2(z^{\mu}(\tau))\right)\Bigg],
\end{aligned}
\end{equation}
which is exactly the form we have introduced in Eq.~\ref{eq:SourceTermIntegration}, and the corresponding virtual amplitude is labeled as $Z^{\infty/\text{H},C}_{lmknj}$. Thus, we have
\begin{equation}\label{eq:fluxInt}
\begin{aligned}
\left<\frac{dC}{d\tilde{\lambda}}\right> = &\lim_{T\rightarrow\infty}\frac{1}{2T}\int_{-T}^T d\tilde{\lambda} \int dw \sum_{lmknj}\frac{1}{2iw^3} \\
&\left(\frac{\tilde{\Gamma}}{2\pi}\overline{Z}^{\infty,C}_{lmknj}Z^{\infty}_{lmw}e^{i\tilde{\Gamma}(w-w_{mknj})\tilde{\lambda}} + \text{the "down" part} + \right) \\
&+ (c.c).
\end{aligned}
\end{equation}
Again, since $Z^{\infty}_{lmw}$ is composed of discrete spectra with $Z^{\infty}_{lmw} = \sum_{k'n'j'}Z^{\infty}_{lmk'n'j'}\delta(w-w_{mk'n'j'})$, the integration in Eq.~\ref{eq:fluxInt} will be finally reduced to
\begin{equation}\label{eq:QC_EquationFlux}
\sum_{lmknj}\frac{\tilde{\Gamma}}{2\pi w_{mknj}^3}\text{Im}\left[\overline{Z}^{\infty,C}_{lmknj}Z^{\infty}_{lmknj} + \alpha_{lmw_{mknj}}\overline{Z}^{\text{H},C}_{lmknj}Z^{\text{H}}_{lmknj} \right].
\end{equation}

Obviously, the above expression holds for all fluxes of $C=\{E,J_z,s_{||},K\}$, differing only in $\frac{\partial C}{\partial v^{\alpha}}$ and $\frac{\partial C}{\partial s^{\alpha\beta}}$.
In the derivation, we did not actually rely on any specific properties of the orbit or the constants, and hence it constitutes a general method for computing the radiative flux for periodic orbits in Kerr spacetime. However, before putting it into practical computation, we must address the question: How can we confirm that the radiative field (half-retarded minus half-advanced field, i.e., $h^\text{rad}_{\mu\nu}$) indeed represents the orbit-averaged part of the effective field (i.e., $h^E_{\mu\nu}$ in the self-force theory)?

For the energy and angular momentum, Ref.~\cite{Akcay2020} rigorously proved this for linearized MPD orbits, which ultimately leads to results consistent with the flux-balance law \cite{Teukolsky1974}.
\begin{equation}\label{eq:FB_EquationFluxE}
\left<\frac{dE}{dt}\right>_\text{FB} = \sum_{lmknj}\frac{1}{4\pi w_{mknj}^2}\left(\lvert Z^{\infty}_{lmknj}\lvert^2 + \alpha_{lmw_{mknj}}\lvert Z^\text{H}_{lmknj}\lvert^2\right),
\end{equation}
\begin{equation}\label{eq:FB_EquationFluxJz}
\left<\frac{dJ_z}{dt}\right>_\text{FB} = \sum_{lmknj}\frac{m}{4\pi w_{mknj}^3}\left(\lvert Z^{\infty}_{lmknj}\lvert^2 + \alpha_{lmw_{mknj}}\lvert Z^\text{H}_{lmknj}\lvert^2\right).
\end{equation}
Therefore, it seems natural to extend it to the fluxes of the other constants of motion, since it is hard to imagine that the properties of $h^E_{\mu\nu}$ would differ for different constants. Or at least, we should regard the flux results from the radiative field as the dominant contribution to the orbit‑averaged evolution. 

Fortunately, before the formal release of this work, Ref.~\cite{Viktor2026}, building upon Ref.~\cite{Josh2025,Grant2025,Witzany2025}, presented a particularly elegant flux-balance formula for the Carter-like constant flux of linearized MPD orbits.
\begin{equation}\label{eq:FB_EquationFluxK}
\begin{aligned}
\left<\frac{dK}{dt}\right>_\text{FB} = &\sum_{lmkn}\Bigg(\left(2 + s_{||}\frac{\tilde{E}}{\sqrt{\tilde{K}}}\right)\left(k\tilde{\Upsilon}_z-w_{mkn}\tilde{\Gamma}_{t,z} + m\tilde{\Upsilon}_{\varphi,z}\right) \\
&-\frac{as_{||}}{\sqrt{\tilde{K}}}\left(m-aw\right)\Bigg)\left(\lvert Z^{\infty}_{lmknj}\lvert^2 + \alpha_{lmw_{mknj}}\lvert Z^\text{H}_{lmknj}\lvert^2\right).
\end{aligned}
\end{equation}
As will be shown by the numerical comparisons in the following sections, our expectation is well confirmed. As for the $s_{||}$, Ref.~\cite{Viktor2024} has already shown that, within the linear-spin approximation, its contribution to its own evolution vanishes. Consequently, orbits for which the orbital angular momentum is aligned with the secondary spin will remain aligned under radiation, and hence we will only consider aligned orbits in the following.

\section{Results}\label{Sec:Result}
For the orbit part, the elliptic functions are computed by the GNU Scientific Library, and the relative error is set as GSL\_PREC\_DOUBLE ($\sim 10^{-16}$). For the perturbation part, the relative error in the Jiang-Han method is set below $10^{-15}$. All the calculations presented in this paper are performed using double-precision floating-point numbers.

\subsection{Code Validation and Consistency Checks}
Before presenting our main results, we perform several checks to validate the numerical implementation of our Teukolsky-based perturbation calculation. The orbital motion of the spinning secondary is handled analytically using the closed-form expressions derived in Ref.~\cite{Skoupy2025}, which have been independently verified. Therefore, our validation focuses exclusively on the perturbation sector—specifically, on the computation of the source term and the resulting asymptotic amplitudes. Three complementary tests are carried out: (i) a comparison with the published energy, angular momentum and Carter constant flux data for generic (eccentric and inclined) orbits in the nonspinning limit in Ref.~\cite{Fujita2009II}; (ii) a comparison with published energy flux data (a mode where $l=2$ and $m=2$) for circular, equatorial orbits with full-spin effects in Ref.~\cite{Piovano2020II}; and (iii) an internal consistency check for generic MPD orbits, where Eq.~\ref{eq:QC_EquationFlux} and the flux-balance formula (i.e., Eq.~\ref{eq:FB_EquationFluxE} and Eq.~\ref{eq:FB_EquationFluxK}) in our code are compared.

\subsubsection{nonspinning limit}
If $s=0$, the result should simply go back to the geodesic case, and the summation of the mode index $j$ disappears. For the geodesic orbit, a more commonly used quantity is $Q$ rather than $K$,
\begin{equation}
Q = K - \left(aE-J_z\right)^2.
\end{equation}
Thus, its orbit-averaged evolution is computed from
\begin{equation}
\left<\frac{dQ}{dt}\right> = \left<\frac{dK}{dt}\right> - 2\left(aE-J_z\right)\left(a\left<\frac{dE}{dt}\right> - \left<\frac{dJ_z}{dt}\right>\right).
\end{equation}

The Fourier expansion in Eq.~\ref{eq:3DFourierExpansion} is still realized by the Trapezoidal Rule with a fixed grid of $100 \times 100/200\times 200$ uniformly spaced points. The truncation of the mode number is also fixed at $\left\{l_{\text{max}} = 10/20, k_{\text{max}} = 10, n_{\text{max}}=10 \right\}$. 

Typically, on a single CPU, 13th Gen Intel(R) Core(TM) i5-13500HX (2.50 GHz), it costs $3.27695\times 10^{2}$ seconds for the calculations of a total of 27018 modes with the $100\times 100$ grid.

We compare the flux-balance data (from Eq.~\ref{eq:FB_EquationFluxE}, Eq.~\ref{eq:FB_EquationFluxJz} and Eq.~\ref{eq:FB_EquationFluxK}) with Ref.~\cite{Fujita2009II} in Tab.~\ref{tab:NonspinningLimit}. In Ref.~\cite{Fujita2009II}, the author fixed $l_{\text{max}}$ to $20$, while adjusted $k_{\text{max}}$, $n_{\text{max}}$ and the grid with the adaptive precision. From Tab.~\ref{tab:NonspinningLimit}, we find that for the near‑equatorial orbit with small eccentricity ($\{\theta_\text{min}=70^{\circ},e=0.1\}$), the settings of $\{l_\text{max}=20,k_\text{max}=10,n_\text{max}=10,100\times 100\}$ already bring the relative difference to $10^{-8}$--$10^{-10}$, and further refining the integration grid to $200\times 200$ yields only marginal improvement. For orbits with larger eccentricity or farther from the equatorial plane, the relative differences are correspondingly larger. This is expected, since in these cases the contributions from higher‑$k$ and higher‑$n$ modes are more significant, and those modes are not included in the calculation.


\begin{table*}
  \centering
  \caption{nonspinning limit for the flux data. For every block, the data in the first line comes from Ref.~\cite{Fujita2009II}, and the others come from our code with the flux-balance formula. In Ref.~\cite{Fujita2009II}, their mode truncation (for $n_{\text{max}}$ and $k_{\text{max}}$) and Fourier expansion adopt the adaptive precision method, so we neglect its $l_{\text{max}}$ and grid parameters in this table, although their $l_{\text{max}}$ is actually set as 20. In the table, $*$ refers to the same data in the previous line. For the geodesic parameter, $p$ is always set as $6M$, and $a$ is set as $0.9M$. Although we haven't defined these root parameters for generic MPD orbits, they can be regarded as the root parameters of the geodesic orbit, which shares the same values of $\{E,J_z,K\}$ with the MPD orbit.}
  \begin{tabular}{|cccc|ccc|}
  \hline
  $e$ & $\theta_{\text{min}}$ & $l_{\text{max}}$ & \text{grid} & $\left<dE/dt\right>^{\infty}_\text{FB}$ & $\left<dJ_z/dt\right>^{\infty}_\text{FB}$ & $\left<dQ/dt\right>^{\infty}_\text{FB}$ \\
  \hline
  0.1 & $70^{\circ}$ & - & - & $-5.8736380008 \times 10^{-4}$ & $-8.5372788158 \times 10^{-3}$ & $-5.2401984854 \times 10^{-3}$ \\
  * & * & 10 & $100\times 100$ & $-5.8736175510 \times 10^{-4}$ & $-8.5372509708 \times 10^{-3}$ & $-5.2401811331 \times 10^{-3}$ \\
  * & * & 10 & $200\times 200$ & * & * & * \\
  * & * & 20 & $100\times 100$ & $-5.8736380025 \times 10^{-4}$ & $-8.5372788186 \times 10^{-3}$ & $-5.2401984858 \times 10^{-3}$ \\
  \hline
  0.1 & $30^{\circ}$ & - & - & $-6.8334819527 \times 10^{-4}$ & $-6.0782911169 \times 10^{-3}$ & $-4.3219465021 \times 10^{-2}$ \\
  * & * & 10 & $100\times 100$ & $-6.8334461815 \times 10^{-4}$ & $-6.0782623431 \times 10^{-3}$ & $-4.3219247641 \times 10^{-2}$ \\
  * & * & 10 & $200\times 200$ & * & * & * \\
  * & * & 20 & $100\times 100$ & $-6.8334812015 \times 10^{-4}$ & $-6.0782912888 \times 10^{-3}$ & $-4.3219457427 \times 10^{-2}$ \\
  \hline
  0.3 & $70^{\circ}$ & - & - & $-6.8040992971 \times 10^{-4}$ & $-8.6259076212 \times 10^{-3}$ & $-5.2214505250 \times 10^{-3}$ \\
  * & * & 10 & $100\times 100$ & $-6.8036654072 \times 10^{-4}$ & $-8.6255146425 \times 10^{-3}$ & $-5.2212280999 \times 10^{-3}$ \\
  * & * & 10 & $200\times 200$ & * & * & * \\
  * & * & 20 & $100\times 100$ & $-6.8037350043 \times 10^{-4}$ & $-8.6255916121 \times 10^{-3}$ & $-5.2212211109 \times 10^{-3}$ \\
  \hline
  0.3 & $30^{\circ}$ & - & - & $-8.3059757658 \times 10^{-4}$ & $-6.4967420401 \times 10^{-3}$ & $-4.5070180371 \times 10^{-2}$ \\
  * & * & 10 & $100\times 100$ & $-8.3042528613 \times 10^{-4}$ & $-6.4956818087 \times 10^{-3}$ & $-4.5063518271 \times 10^{-2}$ \\
  * & * & 10 & $200\times 200$ & * & * & * \\
  * & * & 20 & $100\times 100$ & $-8.3043105122 \times 10^{-4}$ & $-6.4957210867 \times 10^{-3}$ & $-4.5063819644 \times 10^{-2}$ \\
  \hline
  \hline
  $e$ & $\theta_{\text{min}}$ & $l_{\text{max}}$ & \text{grid} & $\left<dE/dt\right>^{\text{H}}_\text{FB}$ & $\left<dJ_z/dt\right>^{\text{H}}_\text{FB}$ & $\left<dQ/dt\right>^{\text{H}}_\text{FB}$ \\
  \hline
  0.1 & $70^{\circ}$ & - & - & $4.2524561258 \times 10^{-6}$ & $6.7150025467 \times 10^{-5}$ & $1.4107369641 \times 10^{-6}$ \\
  * & * & 10 & $100\times 100$ & $4.2524561272 \times 10^{-6}$ & $6.7150025482 \times 10^{-5}$ & $1.4107370208 \times 10^{-6}$ \\
  * & * & 10 & $200\times 200$ & * & * & * \\
  * & * & 20 & $100\times 100$ & * & * & * \\
  \hline
  0.1 & $30^{\circ}$ & - & - & $3.3311347714 \times 10^{-6}$ & $1.1167703064 \times 10^{-4}$ & $-1.0235395371 \times 10^{-4}$ \\
  * & * & 10 & $100\times 100$ & $3.3311348082 \times 10^{-6}$ & $1.1167703176 \times 10^{-4}$ & $-1.0235395458 \times 10^{-4}$ \\
  * & * & 10 & $200\times 200$ & * & * & * \\
  * & * & 20 & $100\times 100$ & * & * & * \\
  \hline
  0.3 & $70^{\circ}$ & - & - & $5.8696781544 \times 10^{-6}$ & $7.7672745798 \times 10^{-5}$ & $-5.8620517437 \times 10^{-6}$ \\
  * & * & 10 & $100\times 100$ & $5.8696762908 \times 10^{-6}$ & $7.7672734849 \times 10^{-5}$ & $-5.8620545901 \times 10^{-6}$ \\
  * & * & 10 & $200\times 200$ & * & * & * \\
  * & * & 20 & $100\times 100$ & * & * & $-5.8620545902 \times 10^{-6}$ \\
  \hline
  0.3 & $30^{\circ}$ & - & - & $5.1949453031 \times 10^{-6}$ & $1.6611683580 \times 10^{-4}$ & $-2.2665864653 \times 10^{-4}$ \\
  * & * & 10 & $100\times 100$ & $5.1949742333 \times 10^{-6}$ & $1.6611680706 \times 10^{-4}$ & $-2.2665716127 \times 10^{-4}$ \\
  * & * & 10 & $200\times 200$ & * & * & * \\
  * & * & 20 & $100\times 100$ & $5.1949742335 \times 10^{-6}$ & $1.6611680707 \times 10^{-4}$ & $-2.2665716128 \times 10^{-4}$ \\
  \hline
  \end{tabular}
  \label{tab:NonspinningLimit}
\end{table*}

\subsubsection{circular orbit}

The circular (equatorial and spin-aligned) case of MPD orbits has been detailedly studied as in Ref.~\cite{Piovano2020II}. We wish to compare our results with theirs, but the orbit in Ref.~\cite{Piovano2020II} is solved by full-spin effects instead of linear-spin approximation. Fortunately, our code adopts a modular design, so it is very convenient to replace the orbit-source part. Thus, the strict circular orbit in this part is kept the same as in Ref.~\cite{Piovano2020II}, instead of the limiting circular case of the generic orbit introduced in Sec.~\ref{Sec:MPDOrbit}.

We present the results in Tab.~\ref{tab:CircularOrbit}. In the case of a circular orbit with a secondary spin, the secondary object should stay in the equatorial plane, and the secondary spin should be aligned to the primary one (i.e., $s_{\perp}=0$). So the mode summation of $\{k,n,j\}$ disappears, and we only consider the mode where $\{l=2,m=2\}$. The normalized 22-mode energy flux, $\hat{\mathcal{F}}^{\infty}_{22}$, is defined as
\begin{equation}
\hat{\mathcal{F}}^{\infty}_{22} := \left<\frac{dE}{dt}\right>^{\infty}_{22}\Big/\left(\frac{32}{5}\vert\hat{\Omega}\vert^{\frac{10}{3}}\right)
\end{equation}
where $32\vert\hat{\Omega}\vert^{\frac{10}{3}}/5$ is the leading post-Newtonian order of the 22-mode energy flux, and $\hat{\Omega}$ is the azimuthal frequency, $\hat{\Omega} = d\varphi/dt$. Note that in order to amplify the effect of the secondary spin, the value of $s$ is extremely and unphysically large, e.g., $\sim 1$. As can be seen from Tab.~\ref{tab:CircularOrbit}, there is a high degree of agreement between the results of the two.
\begin{table}
    \centering
    \caption{the 22-mode data of the normalized energy flux for the circular orbit. "GAP" refers to the data from Ref.~\cite{Piovano2020II}, and "FB" refers to the data from our code with the flux-balance formula. $a=0.9$ refers to the prograde case, while $a=-0.9$ refers to the retrograde case.}
    \begin{tabular}{|ccc|cc|}
    \hline
    $a/M$ & $r/M$ & $s/M$ & $\hat{\mathcal{F}}^{\infty}_{22,\text{GAP}}$ & $\hat{\mathcal{F}}^{\infty}_{22,\text{FB}}$  \\
    \hline
    $0$ & $10$ & $-0.9$ & 0.8779 & $8.7790838643\times 10^{-1}$ \\
    $0$ & $10$ & $0.9$ & 0.7987 & $7.9865623458\times 10^{-1}$ \\
    \hline
    $0.9$ & $10$ & $-0,9$ & 0.7716 & $7.7164706579\times 10^{-1}$ \\
    $0.9$ & $10$ & $0.9$ & 0.7560 & $7.5603115795\times 10^{-1}$ \\
    \hline
    $-0.9$ & $10$ & $-0.9$ & 0.8816 & $8.8159713066\times 10^{-1}$ \\
    $-0.9$ & $10$ & $0.9$ & 1.0519 & $1.0519030350$ \\
    \hline 
    \end{tabular}
    \label{tab:CircularOrbit}
\end{table}

\subsubsection{consistency check}\label{sec:consistency}
The two numerical comparisons with published results presented above are both based on the flux‑balance formula, which verifies that the asymptotic amplitude computation in our code is correct. Finally—and this is the central focus of this paper—we will compare whether the results obtained in our code from the flux‑balance formula (Eq.~\ref{eq:FB_EquationFluxE} and Eq.~\ref{eq:FB_EquationFluxK}) and from the radiation field (Eq.~\ref{eq:QC_EquationFlux}) are consistent with each other.

The results are shown in Tab.~\ref{Tab:fluxMPDaligned}. The behavior of angular-momentum fluxes is quite similar to the energy case; thus, it is not included in the table. Again, $l_\text{max}=20$, $k_\text{max}=n_\text{max}=10$ and the grid is set as $100\times 100$. From this table, we can clearly see the relation that (the relative difference between fluxes from Eq.~\ref{eq:QC_EquationFlux} and from the flux-balance formula) $\Delta_\text{FB} \propto s^2$, which indicates the consistency between Eq.~\ref{eq:QC_EquationFlux} and the flux-balance law as the linear-spin approximation is adopted in this paper.

We are not surprised by this consistency. Indeed, using the properties of the orbit and those of the constants of motion, one should in principle be able to express Eq.~\ref{eq:QC_EquationFlux} in terms of the asymptotic amplitudes, eventually yielding a form identical to that of the flux‑balance law, just as was done in Ref.~\cite{Sago2006} for the geodesic case—except that for spinning particles the procedure becomes more involved. Physically, this consistency implies that, even for spinning particles, the radiative field (half-retarded minus half-advanced field) still governs the orbit‑averaged evolution.

\begin{table*}
  \centering
  \caption{Radiation fluxes from generic MPD orbits with different 2nd spins. "QC" refers to the data computed from Eq.~\ref{eq:QC_EquationFlux}. $a$ is always set as $0.9M$. For $s=0$, the orbit is set as the geodesic where $\{p=6M, e=0.1, \theta_{\text{min}}=70^{\circ}\}$. For $s\neq 0$, the orbit is set as the MPD orbit with $s_{\perp}=0$ and sharing the same values of $\{E,J_z,K\}$ with the $s=0$ case. $\Delta_\text{g}$ is the relative shift between the $s = 0$ case and the $s\neq 0$ case, and $\Delta_\text{FB}$ is the relative difference with the data from the flux-balance formula.}
  \label{Tab:fluxMPDaligned}
  \begin{tabular}{|c|ccc|ccc|}
  \hline
  $s/M$ & $\left<dE/dt\right>^{\infty}_\text{QC}$ & $\Delta_\text{FB}$ & $\Delta_\text{g}$ & $\left<dK/dt\right>^{\infty}_\text{QC}$ & $\Delta_\text{FB}$ & $\Delta_\text{g}$ \\
  \hline
  0 & $-5.8736380025\times 10^{-4}$ & $2.8\times 10^{-15}$ & $0$ & $-3.4384854875\times 10^{-2}$ & $3.6\times 10^{-15}$ & $0$ \\
  \hline
  0.00001 & $-5.8737690863\times 10^{-4}$ & $-1.1\times 10^{-10}$ & $2.2\times 10^{-5}$ & $-3.4385043137\times 10^{-2}$ & $-1.9\times 10^{-10}$ & $5.5\times 10^{-6}$ \\
  \hline
  0.00002 & $-5.8739001643\times 10^{-4}$ & $-4.2\times 10^{-10}$ & $4.5\times 10^{-5}$ & $-3.4385231355\times 10^{-2}$ & $-7.4\times 10^{-10}$ & $1.1\times 10^{-5}$ \\
  \hline
  0.00003 & $-5.8740312564\times 10^{-4}$ & $-9.2\times 10^{-10}$ & $6.7\times 10^{-5}$ & $-3.4385419647\times 10^{-2}$ & $-1.6\times 10^{-9}$ & $1.6\times 10^{-5}$ \\
  \hline
  0.00004 & $-5.8741623463\times 10^{-4}$ & $-1.6\times 10^{-9}$ & $8.9\times 10^{-5}$ & $-3.4385607920\times 10^{-2}$ & $-2.9\times 10^{-9}$ & $2.2\times 10^{-5}$ \\
  \hline
  0.00005 & $-5.8742934373\times 10^{-4}$ & $-2.5\times 10^{-9}$ & $1.1\times 10^{-4}$ & $-3.4385796189\times 10^{-2}$ & $-4.5\times 10^{-9}$ & $2.7\times 10^{-5}$ \\
  \hline
  \hline
  $s/M$ & $\left<dE/dt\right>^\text{H}_\text{QC}$ & $\Delta_\text{FB}$ & $\Delta_\text{g}$ & $\left<dK/dt\right>^\text{H}_\text{QC}$ & $\Delta_\text{FB}$ & $\Delta_\text{g}$ \\
  \hline
  0 & $4.2524561272\times 10^{-6}$ & $2.4\times 10^{-15}$ & $0$ & $2.3185174288\times 10^{-4}$ & $3.7\times 10^{-15}$ & $0$ \\
  \hline
  0.00001 & $4.2526597555\times 10^{-6}$ & $-9.3\times 10^{-11}$ & $4.8\times 10^{-5}$ & $2.3185699280\times 10^{-4}$ & $-1.9\times 10^{-10}$ & $2.3\times 10^{-5}$ \\
  \hline
  0.00002 & $4.2528633820\times 10^{-6}$ & $-3.6\times 10^{-10}$ & $9.6\times 10^{-5}$ & $2.3186224242\times 10^{-4}$ & $-7.4 \times 10^{-10}$ & $4.5\times 10^{-5}$ \\
  \hline
  0.00003 & $4.2530670238\times 10^{-6}$ & $-9.8\times 10^{-10}$ & $1.4\times 10^{-4}$ & $2.3186749268\times 10^{-4}$ & $-1.7\times 10^{-9}$ & $6.8\times 10^{-5}$ \\
  \hline
  0.00004 & $4.2532706660\times 10^{-6}$ & $-1.9\times 10^{-9}$ & $1.9\times 10^{-4}$ & $2.3187274277\times 10^{-4}$ & $-3.0 \times 10^{-9}$ & $9.1\times 10^{-5}$ \\
  \hline
  0.00005 & $4.2534743119\times 10^{-6}$ & $-3.1\times 10^{-9}$ & $2.4\times 10^{-4}$ & $2.3187799281\times 10^{-4}$ & $-4.8\times 10^{-9}$ & $1.1\times 10^{-4}$ \\
  \hline
  \end{tabular}
  \label{tab:GenericOrbits}
\end{table*}

\subsection{Inspiral Waveforms from Generic MPD Orbits}
As a direct application of the fluxes, we present an example of inspiral waveforms with a spinning secondary based on our code. Fig.~\ref{fig:inspiral} shows a segment of inspiral waveforms with the $s=0$ case and the $s\neq 0$ case. The initial $\{E,J_z,K\}$ are set to share the same value with a geodesic of $\{a=0.9M,p=10M,e=0.1,\theta_\text{min}=70^{\circ}\}$. The field point at which the gravitational perturbation is evaluated is set to $\{r\rightarrow\infty,z=1/2,\varphi=\pi/2\}$. Again, the grid is fixed with $100 \times 100$, and the truncation of the mode is fixed at $\left\{l_{\text{max}} = 20, k_{\text{max}} = 10, n_{\text{max}}=10 \right\}$. The inspiral waveforms can be evaluated by
\begin{equation}
h_{+}(t) - ih_{\times}(t) = \sum_{lmkn} \frac{-2}{w^2_{mkn}}\frac{1}{r}Z^{\infty}_{lmkn}e^{im\varphi-i\Phi_{mkn}(t)}\frac{S^{aw_{mkn}}_{lm}}{\sqrt{2\pi}}, 
\end{equation}
where the phase $\Phi_{mkn}(t)$ is defined as
\begin{equation}
\Phi_{mkn}(t) = \int^t_{t_0} w_{mkn}(t)dt.
\end{equation}
When performing the orbital evolution, the mass ratio of the EMRI system is set to $10^{-4}$, so the $s=0.0001M$ case actually corresponds to a situation with an extreme secondary spin.

\begin{figure*}
    \centering
    \includegraphics[width=1.\linewidth]{waveform_inspiral.jpg}
    \caption{a segment of inspiral waveforms. The initial orbit condition (where $t-r_*=0$) is set as $\{a=0.9M,E=0.95286602596,J_z=3.2638133216M,K=7.2098629692M^2\}$ with different secondary spins.}
    \label{fig:inspiral}
\end{figure*}

\section{Summary}
In this work, we calculated the gravitational perturbation induced by the linearized MPD orbit, and then used the radiative field to derive the evolution of the MPD orbit---namely, the evolution equations for its constants of motion. The core method employed for this derivation is presented in Sec.~\ref{Sec:OrbitInspiral}, and the final result is Eq.~\ref{eq:QC_EquationFlux}. We then verify its correctness in Sec.~\ref{sec:consistency} by comparing it with the results from the flux‑balance law.

Regarding Eq.~\ref{eq:QC_EquationFlux}, we should recognize that it is indeed less convenient to use than the flux‑balance formula, especially because it requires the computation of additional virtual amplitudes, whereas the flux‑balance law only needs the asymptotic amplitudes. Nevertheless, Eq.~\ref{eq:QC_EquationFlux} provides a numerical but generalized procedure for incorporating the radiative field into the orbital evolution framework for periodic orbits in Kerr spacetime, without relying on complicated algebraic simplifications. At the same time, as we emphasized earlier, before applying the radiative field to orbital evolution, one must carefully consider its role in the self-force calculation.

There are several aspects of this work that can be further optimized. For instance, extracting the spin corrections from the flux calculations separately would greatly facilitate the future integration of spin corrections with other 1PA effects. In addition, all calculations in this work are currently performed on CPUs. Migrating the code to GPUs would greatly enhance computational efficiency, as done in the FastEMRIWaveforms project \cite{FEW}.

\begin{acknowledgments}
This work is supported by the National Key R\&D Program of China (Grant No. 2021YFC2203002), NSFC (National Natural Science Foundation of China) No. 12473075 and No. 12173071, National Science and Technology Major Project (No. 2024ZD1100601). This work made use of the High Performance Computing Resource in the Core Facility for Advanced Research Computing at Shanghai Astronomical Observatory.
\end{acknowledgments}

\newpage

\appendix

\bibliography{main}
\end{document}